\documentclass[a4paper, 11pt]{article}
\usepackage[pdftex, a4paper]{geometry}
\usepackage{graphicx}				
				
\usepackage{fancyhdr}
\usepackage{amsmath}
\usepackage{amsfonts}
\usepackage{dsfont}
\usepackage{mathrsfs}
\usepackage{amssymb}
\usepackage{bbm}
\usepackage{epsfig}
\usepackage[english]{babel}
\usepackage[all]{xy}

\newtheorem{theorem}{Theorem}
\newtheorem{definition}{Definition}

\newtheorem{lemma}{Lemma}
\newtheorem{remark}{Remark}

\begin{document}

\title{\bf Reaching a Quantum Consensus: Master Equations that Generate  Symmetrization and Synchronization}
\date{}

\author{Guodong Shi, Daoyi Dong, Ian R. Petersen,\\ and Karl Henrik Johansson}
\maketitle

\begin{abstract}
In this paper, we propose and study a master-equation based approach to drive   a quantum network with $n$ qubits
to a consensus (symmetric) state introduced by Mazzarella et al. The state evolution of the quantum network is  described by a Lindblad master equation with  the Lindblad terms
generated by  continuous-time swapping operators,  which also introduce
  an underlying interaction graph.  We establish a graphical method that bridges   the proposed quantum consensus scheme and  classical consensus dynamics by studying an induced graph
      (with $2^{2n}$ nodes) of the  quantum interaction graph (with $n$ qubits). A fundamental connection  is then shown  that quantum consensus over the quantum graph
      is equivalent to componentwise classical consensus over the induced graph, which allows various existing works on classical consensus to be  applicable to the quantum setting. Some basic scaling and structural  properties of the quantum induced graph are established via combinatorial analysis. Necessary and sufficient conditions for exponential and  asymptotic quantum consensus are obtained, respectively,
      for switching quantum interaction graphs. As a quantum analogue of classical synchronization of coupled oscillators, quantum synchronization conditions
      are also presented, in which the reduced states of all qubits tend to a common trajectory.
\end{abstract}


\section{Introduction}
In the past decades, distributed control and optimization methods have witnessed
a wide range of applications in  network systems such as multi-vehicle systems,
wireless communication networks, smart grids,  and social networks \cite{Magnus}-\cite{mor}.
 A networked system consists of a number of interconnected nodes, often denoted agents,
 each of which represents an individual functioning unit ranging from  a robot, a power generator,
 to a member of a society.   Recent development in quantum physics and quantum information science suggests
 the possibility of modeling and analyzing  quantum systems as networks of quantum nodes \cite{Nielsen}-\cite{Gough and James 2009}.
 In these networks,  each quantum node (agent) represents a photon, an electron, an atom, or a finite dimensional quantum system.
    Nodes in a quantum network are described by quantum mechanics and the interactions between different agents involve non-classical correlations.
    These unique quantum characteristics make the development of distributed solutions   in quantum networks more difficult than  classical network systems  \cite{Nielsen}.
    It is interesting to  understand how synchronization and consensus in quantum networks relate to traditional networks, and if the wealth of  graph-theoretic   tools
    recently developed for traditional networks are also applicable  to quantum networks.

One of the primary  objectives in distributed control and coordination is to drive a network to a consensus,
where all agents hold the same state, by local interactions  \cite{saber04}-\cite{shisiam}.
Several efforts have been made to investigate the consensus problem in the quantum domain under
discrete-time settings. Sepulchre \emph{et al.}  \cite{Sepulchre-non-communtative} generalized consensus algorithms to non-commutative spaces
 and presented convergence results for quantum stochastic maps. They showed how the Birkhoff theorem can be used to analyze the asymptotic convergence
 of a quantum system to a fully mixed state.  Mazzarella \emph{et al.} \cite{Ticozzi} made a systematic study regarding   consensus-seeking in quantum networks,
 and pointed out that consensus in a quantum network has close connection to distributed quantum computation, quantum communication and quantum random walk.
 Four classes of consensus quantum states based on invariance and symmetry properties were introduced, and a quantum generalization to the gossip iteration
 algorithm (e.g., \cite{Boyd-gossip}) was proposed for reaching a symmetric state (consensus) over a  quantum network. Such a quantum gossip iteration algorithm
 is realized through discrete-time quantum swapping operations between two subsystems in a quantum network and can make the quantum network converge to symmetric
 states while preserving the expected values  of permutation-invariant global observables. The class of quantum gossip algorithms can be further extended to
 so-called symmetrization problems in a group-theoretic framework and be applied to consensus on probability distributions and quantum dynamical decoupling \cite{Ticozzi-SIAM}.

Quantum systems with external inputs are modeled using master equations that define continuous-time quantum state evolution \cite{Ticozzi-Viola-2008}-\cite{Dong 2013TAC}.
One of the
simplest cases is when a Markovian approximation can be applied
under the assumption of a short environmental correlation time permitting the neglect of memory
effects \cite{Breuer and Petruccione 2002}, where a so-called  Lindblad equation can be employed to describe
the quantum state evolution. In this paper, we show that a Lindblad master equation \cite{Breuer and Petruccione 2002}, \cite{Lindblad 1976}
can be obtained with  the Lindblad terms generated by  swapping operators among the qubits, for the dynamical evolution of the quantum network.
 The swapping operations also introduce   an underlying interaction graph for the quantum network, which indeed leads to a distributed structure  for the master equation.
  In this way, a continuous-time generalization of  the work of \cite{Ticozzi,Ticozzi-SIAM} is introduced, under legitimate quantum state
   evolution\footnote{The continuous-time generalization of \cite{Ticozzi,Ticozzi-SIAM}  for quantum consensus with
   fixed but general quantum permutation interactions,
   was also independently presented in \cite{Ticozzi-MTNS}, where a necessary and sufficient condition
    was derived for reaching quantum symmetric consensus from a  group-theoretical point of view.}.

The contributions of the current paper are highlighted as follows.
\begin{itemize}
  \item   A graphical approach  is established  bridging   the proposed quantum consensus scheme and  classical consensus dynamics by introducing an induced graph
      (with $2^{2n}$ nodes) of the  quantum interaction graph (with $n$ qubits). A fundamental connection  is shown  that quantum consensus evolution over the $n$-qubit network is equivalent to a number of independent classical consensus processes,
      running in parallel over the connected components    of the induced graph. Several fundamental scaling and structural properties are obtained for the  induced graph.
  The number of components is  characterized;
     tight bounds of component sizes and node degrees are explicitly given; the induced graph is shown to be regular and the diagonal induced graph is proved to be almost strongly
     regular.

     \item  The  graphical  approach provides  a powerful tool in studying quantum network dynamics via their classical counterparts. Making use of existing understandings
     of classical consensus, we show how to carry out convergence speed optimization via convex programming. We also establish  two necessary
      and sufficient conditions for exponential and asymptotic  quantum consensus, respectively, for switching quantum interaction graphs.

         \item The possibility of quantum synchronization is also investigated, in the sense that the  trajectory  of each qubit (given  by
         the reduced state under partial trace with respect to the space of other qubits) tends to the same trajectory. We show that quantum
         synchronization can be achieved if the
         network Hamiltonian admits an exact tensor product form (or Kronecker sum form) of identical Hamiltonians for each qubit. The trajectory
         synchronization of qubits serves as the quantum analogue of classical synchronization \cite{WU-CHUA95,PRL1998}.
\end{itemize}

The developments of the above quantum consensus  results are inspired and   heavily rely on the concepts  introduced in   \cite{Ticozzi}.  We  study qubit networks for the ease of presentation. Generalization to  network of quantum nodes with identical but greater than two dimensional
Hilbert spaces is straightforward.
We remark that the proposed graphical approach applies directly also to the discrete-time quantum consensus dynamics \cite{Ticozzi}, and thus the corresponding
convergence rate characterization  and optimization can be obtained  using the results in \cite{Boyd-gossip}. We believe that our results add to the  understanding of distributed control and state manipulation of quantum networks.
The graphical approach proposed in the  paper can also be  useful for a larger class of quantum network control problems.

This rest of the paper is organized as follows. Section \ref{Sec2} presents some preliminaries including relevant concepts  in linear algebra, graph theory
and quantum systems. The $n$-qubit network model and its state evolution  master equations are presented in Section~\ref{Secproblem}.
Section \ref{SecGraphical}
 is devoted to a systematic study of the relation between a quantum interaction graph and its induced graph.
 Section \ref{Secsynchronization} establishes quantum synchronization conditions making use of the graphical approach. Section \ref{Sec6} concludes this paper with a few remarks.

\section{Preliminaries}\label{Sec2}
In this section, we introduce some concepts and theory from  linear algebra \cite{Horn}, graph theory \cite{godsil}, and quantum systems \cite{Nielsen}.
\subsection{Matrix Vectorization and Ger\v{s}gorin Theorem}
Given a matrix $M\in \mathbb{C}^{m\times n}$, the vectorization of $M$, denoted by ${\rm \bf vec}(M)$,
is the $mn\times 1$ column vector  $([M]_{11}, \dots,  [M]_{m1},  \dots, [M]_{1n},\dots, [M]_{mn})^T$.
We have ${\rm \bf vec}(ABC)=(C^T\otimes A){\rm \bf vec}(B)$ for all  matrices $A,B,C$ with $ABC$ well defined, where $\otimes$
stands for the Kronecker product. We always use $I_\ell$ to denote the $\ell\times \ell$ identity matrix, and $\mathbf{1}_\ell$ for the all one vector in $\mathbb{R}^\ell$.

The following is the Ger\v{s}gorin disc Theorem which will be used in the proof of main results.

\begin{lemma}\label{lem1} (pp. 344, \cite{Horn})
Let $A=[a_{jk}]\in \mathbb{C}^{n\times n}$. Then all eigenvalues of $A$ are located in the union of $n$ discs
$$
\bigcup_{i=1}^n \Big\{z\in\mathbb{C}: |z-a_{ii}|\leq \sum_{j=1,j\neq i}^n |a_{ij}| \Big\}.
$$
\end{lemma}

\subsection{Graph Theory Essentials}
A simple undirected graph  $\mathrm
{G}=(\mathrm {V}, \mathrm {E})$ consists of a finite set
$\mathrm{V}=\{1,\dots,N\}$ of nodes and an edge set
$\mathrm {E}$, where  an element $e=\{i,j\}\in\mathrm {E}$ denotes   an
{\it edge}  between two distinct nodes $i\in \mathrm{V}$  and $j\in\mathrm{V}$. Two nodes $i,j\in\mathrm{E}$ are said to be {\it adjacent}
if $\{i,j\}$ is an edge  in $\mathrm{E}$. The number of adjacent nodes  of $v$ is called its degree, denoted ${\rm deg}(v)$.
The nodes that are adjacent with a node $v$ as well as itself are called its neighbors. A graph  $\mathrm
{G}$ is called to be {\it regular} if all the nodes have the same degree. A path between two vertices $v_1$ and $v_k$ in $\mathrm{G}$ is a sequence of distinct nodes
$
v_1v_2\dots v_{k}$
such that for any $m=1,\dots,k-1$, there is an edge between $v_m$ and $v_{m+1}$. A pair of distinct  nodes $i$ and $j$
is called to be {\it reachable} from each other if there is a path between them.  A node is always assumed to be reachable
from itself. We call graph $\mathrm{G}$ {\it connected} if every pair of distinct nodes in $\mathrm{V}$ are reachable from each other.
 A  subgraph of $\mathrm{G}$ associated with node set $\mathrm{V}^\ast \subseteq \mathrm{V}$, denoted as $\mathrm{G}|_{\mathrm{V}^\ast}$,
 is the graph $(\mathrm{V}^\ast, \mathrm{E}^\ast)$, where $\{i,j\}\in \mathrm{E}^\ast$ if and only if $\{i,j\}\in \mathrm{E}$ for $i,j\in \mathrm{V}^\ast$.
   A connected component (or just component) of  $\mathrm
{G}$ is a connected subgraph induced by some $\mathrm{V}^\ast \subseteq \mathrm{V}$, which is connected to no additional nodes in $\mathrm
{V}\setminus \mathrm{V}^\ast$.

The (weighted) Laplacian of $\mathrm
{G}$, denoted $L(\mathrm
{G})$, is defined as
$$
L(\mathrm
{G})=D(\mathrm
{G})-A(\mathrm
{G}),
$$
where $A(\mathrm{G})$ is the  $N\times N$ matrix given by $[A(\mathrm{G})]_{kj}=[A(\mathrm{G})]_{jk}=
a_{kj}$ for some $a_{kj}>0$ if $\{k,j\} \in \mathrm{E}$ and  $[A(\mathrm{G})]_{kj}=0$ otherwise, and $D(\mathrm
{G})={\rm diag}(d_1,\dots,d_N)$ with $d_k=\sum_{j=1,j\neq k}^N [A(\mathrm{G})]_{kj}$. It is well known that $L(\mathrm
{G})$ is always positive semi-definite, and the following relation holds: \begin{align}\label{numbercomponents}
{\rm rank}(L(\mathrm
{G}))=N-C_\ast(\mathrm{G}),
\end{align} where $C_\ast(\mathrm{G})$ denotes the number of connected  components of $\mathrm{G}$.

\subsection{Quantum Systems}
\subsubsection{Quantum Systems and the Master Equation}
 The state space associated with any isolated quantum system is a complex vector space with inner product, {i.e.}, a Hilbert space.
  The system is completely described by its state vector, which is a unit vector in the system's state space.    The state space of a composite quantum system is the tensor product of the state space of each component system. For an open quantum system, its state can be described by a positive Hermitian density operator (or density matrix) $\rho$ satisfying $\text{tr}(\rho)=1$. In many
situations, a master equation for the evolution of $\rho(t)$ is a suitable
way to describe the dynamics of an open quantum system. One of the
simplest cases is when a Markovian approximation can be applied
under the assumption of a short environmental correlation time permitting the neglect of memory
effects \cite{Breuer and Petruccione 2002}. Markovian master equations
have been widely used to model quantum systems with external inputs in quantum control \cite{Ticozzi-Viola-2008}-\cite{Dong 2013TAC},
especially for Markovian quantum feedback \cite{Wiseman and Milburn 2009}.  Markovian master equations in the Lindblad form are described as
 \cite{Lindblad 1976,Wiseman and Milburn 2009}
\begin{equation}
\dot{\rho}(t)=-\frac{\imath}{\hbar}[H,\rho(t)]+\sum_{k}\gamma_{k} \mathfrak{D}[L_{k}]\rho(t),
\end{equation}
where $H$ is the effective Hamiltonian as a Hermitian operator over the underlying Hilbert space, $\imath^2=-1$, $\hbar$ is the reduced Planck constant,
 the non-negative coefficients $\gamma_{k}$
specify  the relevant relaxation rates, and $$\mathfrak{D}[L_{k}]\rho=L_{k}\rho L_{k}^{\dag}-\frac{1}{2}L_{k}^{\dag}L_{k}\rho-\frac{1}{2}\rho L_{k}^{\dag} L_{k}.$$

\subsubsection{Swapping Operators}
In quantum  systems, the two-dimensional Hilbert space  forms the state-space of {\it qubits} (the most basic quantum system). For any Hilbert space $\mathcal{H}_\ast$,
  it is convenient to use $|\cdot \rangle$, known as the Dirac notion, to denote  a unit (column)  vector in $\mathcal{H}_\ast$  \cite{Nielsen}. Moreover, $|\xi \rangle^\dag$,
   i.e., the complex conjugate transpose  of $|\xi \rangle$, is denoted as $\langle \xi|$.

Let $\mathcal{H}$ be a two-dimensional Hilbert space for qubits.  The standard  computational basis of $\mathcal{H}$ is denoted  by $|0\rangle$ and $|1\rangle$.
 An $n$-qubits quantum network is the composite quantum system of $n$ qubits in the set $\mathsf{V}=\{1,\dots,n\}$, whose state space is
  the Hilbert space $\mathcal{H}^{\otimes n}=\mathcal{H}\otimes \dots \otimes \mathcal{H}$, where $\otimes$ denotes the tensor product.
   The {\it swapping operator} between qubits $i$ and $j$, denoted as $U_{ij}$,  is defined by
\begin{align*}
&{U_{ij}} \big(|q_{1}\rangle \otimes \dots \otimes| q_{i}\rangle \otimes \dots \otimes |q_j\rangle \otimes \dots \otimes |q_n\rangle \big)= \nonumber\\
 &|q_{1}\rangle \otimes \dots \otimes |q_{j}\rangle\otimes \dots \otimes |q_i\rangle\otimes \dots \otimes |q_n\rangle,
\end{align*}
for all $q_i\in \{0,1\}, i=1,\dots,n$. In other words, the swapping operator $U_{ij}$ switches the information held in qubits $i$ and $j$
without changing the states of other qubits.

Additionally, for any $|p\rangle, |q\rangle \in \mathcal{H}_\ast$, we use the notation $|p\rangle \langle q|$ to denote the operator over $\mathcal{H}_\ast$ defined by
$$
\big(|p\rangle \langle q| \big) |\eta\rangle= \Big \langle  |q\rangle,  |\eta\rangle\Big\rangle |p\rangle, \ \ \forall |\eta\rangle \in \mathcal{H}_\ast,
$$
where $\Big\langle \cdot, \cdot \Big\rangle$ represents the inner product that the Hilbert space   $\mathcal{H}_\ast$ is equipped with.
In standard quantum mechanical notation,  the inner product $\Big \langle  |p\rangle,  |q\rangle\Big\rangle $ is denoted as $\big\langle p \big|q \big\rangle$.
\subsubsection{Partial Trace}
Let $\mathcal{H}_A$ and $\mathcal{H}_B$ be the state spaces of  two quantum systems $A$ and $B$, respectively. Their composite system is described by
a density operator $\rho^{AB}$. Let $\mathfrak{L}_A$, $\mathfrak{L}_B$, and  $\mathfrak{L}_{AB}$ be the spaces of (linear) operators over  $\mathcal{H}_A$,
 $\mathcal{H}_B$, and  $\mathcal{H}_A\otimes\mathcal{H}_B$, respectively.   Then the partial trace over system $B$, denoted by ${\rm Tr}_{\mathcal{H}_B}$,
 is an operator mapping  $\mathfrak{L}_{AB}$ to $\mathfrak{L}_{A}$ defined by
$$
{\rm Tr}_{\mathcal{H}_B}\Big(|p_A\rangle  \langle q_A| \otimes  |p_B\rangle  \langle q_B| \Big)= |p_A\rangle  \langle q_A|  {\rm Tr} \Big(  |p_B\rangle  \langle q_B| \Big)
$$
for all $|p_A\rangle, |q_A \rangle\in \mathcal{H}_A, |p_B\rangle, |q_B\rangle \in \mathcal{H}_B$.

The reduced density operator (state) for system $A$, when the composite system is in the state  $\rho^{AB}$, is defined as $\rho^A= {\rm Tr}_{\mathcal{H}_B}(\rho^{AB})$.
The physical interpretation of $\rho^A$ is that $\rho^A$ holds the full information of system $A$ in $\rho^{AB}$. For a detailed introduction, we  refer to \cite{Nielsen}.

\section{Quantum Consensus and Synchronization  Master Equations}\label{Secproblem}
\subsection{Quantum Networks and Interaction Graphs}
Consider  a quantum network  with $n$ qubits. The qubits are indexed in the set $\mathsf{V}=\{1,\dots,n\}$  and  the state space of this $n$-qubit quantum network
 is denoted as the Hilbert space $\mathcal{H}^{\otimes n}=\mathcal{H}\otimes \dots \otimes \mathcal{H}$, where  $\mathcal{H}$  denotes a two-dimensional
  Hilbert space over $\mathbb{C}$. The density operator of  the $n$-qubit network is denoted as  $\rho$.

We define a quantum interaction graph over the $n$-qubit  network as an undirected graph $\mathsf{G}=(\mathsf{V}, \mathsf{E})$, where  each element in $\mathsf{E}$,
called a quantum edge,  is an unordered pair of two distinct qubits  denoted as $\{i,j\}\in\mathsf{E}$ with $i,j\in\mathsf{V}$.  Let $\Omega$ denote the set of all
quantum interaction graphs over node set $\mathsf{V}=\{1,\dots,n\}$. Let $\sigma(\cdot): [0,\infty)\mapsto \Omega$ be a piecewise constant function.
The obtained time-varying graph is then denoted as  $\mathsf{G}_{\sigma(t)}=(\mathsf{V},\mathsf{E}_{\sigma(t)})$. We assume that there is a
constant $\tau_D>0$ as a lower bound between any two consecutive switching instants of $\sigma(t)$.

\subsection{Dynamics}
Let $H$  be the (time-invariant) Hamiltonian of the $n$-qubit quantum network. In this paper, we propose and investigate  the state evolution of the quantum network
 described by the following master equation:
\begin{align}\label{sysLind}
\frac{d \rho}{dt}=-\frac{\imath}{\hbar}[H, \rho]+\sum_{\{j,k\}\in \mathsf{E}_{\sigma(t)}} \alpha_{jk} \Big(U_{jk}\rho U_{jk}^\dag  -\rho\Big),
\end{align}
where $\alpha_{jk}>0$ is a constant marking the weight of edge $\{j,k\}$, and $U_{jk}$ is the swapping operator between $j$ and $k$.

The system (\ref{sysLind}) will be referred to as the {\it quantum synchronization master equation}. When we assume  $H=0$, the system (\ref{sysLind}) is reduced to
\begin{align}\label{sys10}
\frac{d \rho}{dt}=\sum_{\{j,k\}\in \mathsf{E}_{\sigma(t)}} \alpha_{jk} \Big(U_{jk}\rho U_{jk}^\dag  -\rho\Big).
\end{align}
We call the system (\ref{sys10})  the {\it quantum consensus master equation}.

\begin{remark}
The Lindblad evolution (\ref{sysLind}) is a continuous-time analogue of the quantum gossip  algorithm proposed in  \cite{Ticozzi}.
 This continuous-time generalization to the discrete-time dynamics \cite{Ticozzi,Ticozzi-SIAM} has also been independently   investigated
  in \cite{WCICA,Ticozzi-MTNS}. Compared to the results and analysis methodologies in  \cite{Ticozzi,Ticozzi-SIAM, Ticozzi-MTNS}, in this work we provide
  a new approach to investigate the connection between the proposed quantum consensus scheme and  classical consensus dynamics.
  As will be shown in the following discussions, once this connection has been made clear, various results for classical consensus can then be adapted
   to establish convergence conditions  under more  relaxed conditions imposed on quantum interaction graphs.
\end{remark}

\begin{remark}
The system (\ref{sysLind}) is related to the proposed  realization of  $n$-qubit quantum circuits by nearest-neighbor operations  in \cite{NaturePhysiscs}, which showed that the ability to apply arbitrary Lindblad operators implies encoding of quantum circuits with polynomial overhead.  In the system (\ref{sysLind}), the swapping operator $U_{jk}$
represents external interactions between qubit $j$ and qubit $k$ through their local environment (cf., Figure 1 in \cite{NaturePhysiscs}), and  the network Hamiltonian generates  internal qubit interactions.
\end{remark}

\subsection{Objectives}
 A permutation of the set $\mathsf{V}=\{1,\dots,n\}$ is a bijective map from $\mathsf{V}$ onto itself. We denote by $\pi$ such a permutation.
 Particularly, a permutation $\pi$ is called a swapping between $j$ and $k$, denoted by $\pi_{jk}$,  if  $\pi(j)=k$, $\pi(k)=j$, and $\pi(s)=s, s\in \mathsf{V}\setminus\{j,k\}$.
 The set of all permutations of $\mathsf{V}$ forms a group, called the $n$'th permutation group and denoted by $\mathbf{P}=\{\pi\}$. There are $n!$ elements in $\mathbf{P}$.
Given $\pi\in \mathbf{P}$, we define a unitary operator, $U_\pi$, over $\mathcal{H}^{\otimes n}$, by
$$
U_\pi \big(|q_1\rangle  \otimes \dots \otimes |q_n\rangle \big)= |q_{\pi(1)}\rangle \otimes \dots \otimes |q_{\pi(n)}\rangle,
$$
where $q_i=0$ or $1$ for all $i=1,\dots,n$.  Define an operator over the density operators of $\mathcal{H}^{\otimes n}$, $\mathscr{P}_\ast$, by
\begin{align}
\label{quantumconsensus}
\mathscr{P}_\ast (\rho)= \frac{1}{n!} \sum_{\pi \in \mathbf{P}} U_\pi \rho U^\dag_\pi.
\end{align}
Introduced in \cite{Ticozzi},  $\mathscr{P}_\ast (\rho)$ serves as the quantum average of the
$n$-qubit network at the state $\rho$.

Let the initial time be $t_0\geq0$ and let  $\rho(t_0)$ be the initial density operator of the quantum network.  We make the following definition.

\medskip

\begin{definition}
(i) The system (\ref{sys10}) reaches an asymptotic (symmetric-state)  quantum consensus for initial time $t_0\geq0$ and initial state  $\rho(t_0)$ if $
\lim_{t\to \infty} \rho(t)=\mathscr{P}_\ast (\rho(t_0))$.

(ii) The system (\ref{sys10}) reaches global asymptotic (symmetric-state)  quantum  consensus if quantum consensus is achieved for all  $t_0\geq0$ and all initial density operators $\rho(t_0)$.

(iii) The system (\ref{sys10}) reaches  global  exponential (symmetric-state)  quantum consensus, if there exist $ C(\rho(t_0))>0$
(which may depend on the initial state $\rho(t_0)$) and $\gamma>0$ (which does not depend on $\rho(t_0)$) such that
$$
\Big\|\rho(t)- \mathscr{P}_\ast (\rho(t_0))\Big\| \leq C(\rho(t_0))e^{-\gamma (t-t_0)},\ \ t\geq t_0
$$
for all initial times $t_0\geq 0$ and initial states $\rho(t_0)$.
\end{definition}

\medskip

Let $$
\rho^k(t):= {\rm Tr}_{\otimes_{j\neq k} \mathcal{H}_j } \big(\rho(t)\big)
$$
be  the reduced  state of qubit $k$  at time $t$, $k=1,\dots,n$, defined  by the partial trace over the remaining  $n-1$ qubits' space $\otimes_{j\neq k} \mathcal{H}_j$.
 Here $\mathcal{H}_j$ denotes the two-dimensional Hilbert space corresponding to qubit $j$, $j\in\mathsf{V}$.  Note that $\rho^k(t)$ contains all the information that qubit $k$
 holds in the composite state $\rho(t)$. Consistent with the classical definition of complex network synchronization \cite{WU-CHUA95,PRL1998}, we also introduce the following definition for
 quantum (reduced-state) synchronization.

 \medskip

\begin{definition}
(i). The system (\ref{sysLind}) achieves global asymptotic  quantum (reduced-state) synchronization if
\begin{align}
\lim_{t\to \infty} \Big(\rho^k(t)-\rho^m(t)\Big)=0, \ k,m\in\mathsf{V}
\end{align}
for all initial times $t_0$ and initial values $\rho(t_0)$.

(ii). The system (\ref{sysLind}) achieves global exponential quantum (reduced-state)  synchronization if there are two constants $C(\rho(t_0))>0$ and $\gamma>0$ such that
\begin{align}
\Big\|\rho^k(t)-\rho^m(t)\Big\| \leq C(\rho(t_0))e^{-\gamma (t-t_0)},\ \ t\geq t_0
\end{align}
for all $k,m\in\mathsf{V}$.
\end{definition}
\medskip

Note that along the Lindblad master equation (\ref{sys10}),
$\rho(t)$ will be preserved as  positive,  Hermitian, and with trace one,
as long as $\rho(0)$ defines a proper density operator. While the convergence conditions to be derived
 in the paper do not depend on   these properties held by the density operators. Therefore, throughout the rest of the paper,
 we assume that $\rho(t)$ lies in the general space $\mathbb{C}^{ 2^n \times 2^n}$.

%
%
%
%
%
%

\section{The Quantum Laplacian  and  Induced Graph}\label{SecGraphical}

In this section, we explore the connection between the quantum consensus dynamics  (\ref{sys10}) and its classical analogue through  an
induced (classical) graph from a graphical point of view. We introduce the quantum Laplacian matrix associated with a quantum interaction graph
and show that the convergence to quantum consensus is fully governed by this quantum Laplacian.
This inspired us to introduce the induced graph of the quantum interaction graph, and then equivalence is proved between quantum consensus over the interaction graph
and classical consensus over the induced graph.  We also  establish some basic scaling and structural  properties of the induced graph.
\subsection{The Quantum Laplacian}

We introduce   quantum Laplacian associated with the interaction graph $\mathsf{G}$ as follows.

\medskip

\begin{definition}
Let $\mathsf{G}=(\mathsf{V}, \mathsf{E})$ be a quantum interaction graph. The quantum (non-weighted) Laplacian  of $\mathsf{G}$ is defined as $
L_\mathsf{G}:=\sum_{\{j,k\}\in \mathsf{E}} \big(I_{2^n}\otimes I_{2^n}-  U_{jk} \otimes U_{jk }  \big)$.
\end{definition}

\medskip

Some properties of the quantum Laplacian can be clearly observed: ${L}_\mathsf{G}$ is real and symmetric, ${L}_\mathsf{G} \mathbf{1}_{2^{2n}}=0$, and
all the off-diagonal entries of ${L}_\mathsf{G}$ are non-negative. Consequently,
invoking the  Ger\v{s}gorin disc theorem (cf., Lemma~\ref{lem1}) we know that  all nonzero eigenvalues of ${L}_\mathsf{G}$ are positive, and we denote   the smallest eigenvalue other than zero of ${L}_\mathsf{G}$ as $\lambda_2({L}_\mathsf{G})$.

Consider  the following quantum consensus master equation defined over the quantum interaction graph $\mathsf{G}$:
\begin{align}\label{r4}
 \frac{d}{dt} \rho(t)=\sum_{\{j,k\}\in \mathsf{E}} \big(U_{jk}\rho(t) U_{jk}^\dag  -\rho\big).
\end{align}
Then (\ref{r4}) can be exactly written as
\begin{align}\label{r5}
  \frac{d }{dt} {\bf vec}(\rho)=-L_{\mathsf{G}} {\bf vec}(\rho)
\end{align}
under the vectorization $\rho(t)$.

There holds for the system (\ref{r5}) that  ${\bf vec}({\rho}(t))$ converges to a fixed point in the null space of ${L}_\mathsf{G}$ exponentially,
with the convergence speed  given by $\lambda_2({L}_\mathsf{G})$. Moreover,  different from classical definition of the Laplacian,
the multiplicity of the zero eigenvalue of ${L}_\mathsf{G}$ is no longer one, even when the interaction graph $\mathsf{G}$ is connected.
The following lemma provides a characterization of the null space of the quantum Laplacian.
\begin{lemma}\label{lem0}
${\rm ker} ({L}_\mathsf{G})= \big\{{\rm \bf vec}(z):  \mathscr{P}_\ast (z)  =z\big\}$ if $\mathsf{G}$ is connected.
\end{lemma}

The proof of Lemma \ref{lem0} can be found in Appendix A. In light of Lemma \ref{lem0}, it can be easily deduced that the system (\ref{r4}) reaches exponential  quantum consensus as long as $\mathsf{G}$
 is connected, with convergence rate $\lambda_2({L}_\mathsf{G})$. This is consistent with the  results in \cite{Ticozzi,Ticozzi-MTNS}.

\subsection{The Induced Graph}
For  further investigations of  the quantum Laplacian, we introduce  the following definition.

\medskip

\begin{definition} \label{definitioninduce}
The induced graph of the quantum interaction graph $\mathsf{G}$, denoted by $\mathcal{G}=(\mathcal{V},\mathcal{E})$, is defined in that
$\mathcal{V}=\{1,\dots,2^{2n}\}$ and $\{r,s\}\in\mathcal{E}, r\neq s\in \mathcal{V}$ if and only if $\big[L_\mathsf{G}\big]_{rs}\neq 0$.
\end{definition}

\medskip

Making use of Eq. (\ref{numbercomponents}) and noticing that $L_\mathsf{G}$ is the classical Laplacian of the induced graph $\mathcal{G}$,
the following lemma follows from Lemma \ref{lem0} as a  preliminary property between a quantum interaction graph and its induced graph.

\medskip

\begin{lemma}\label{lemnumber}
If the quantum interaction graph $\mathsf{G}$ is connected, then its induced graph $\mathcal{G}$ has exactly $$
{\rm dim} \Big( \big\{{\rm \bf vec}(z):  \mathscr{P}_\ast (z)  =z, z\in \mathbb{C}^{ 2^n \times 2^n} \big\} \Big)={\rm dim} \Big( {\rm ker} (L_\mathsf{G}) \Big)
$$ connected components.
\end{lemma}

\medskip

We let $X(t)=(x_1 (t) \dots x_{4^n}(t))^T:= {\bf vec}(\rho(t))$ so that the system~(\ref{r5}) defines   classical consensus dynamics over the
 induced graph $\mathcal{G}=(\mathcal{V}, \mathcal{E})$ (cf., \cite{jad03,saber04}), where $x_i(t) \in\mathbb{C}$ stands for the state of node
 $i\in\mathcal{V}$ at time $t$. Let the initial time be $t_0=0$. We make the following definition.

 \medskip

\begin{definition}
Componentwise  consensus over the graph $\mathcal{G}$ in the classical sense is achieved for the system~(\ref{r5}) if
$$
\lim_{t\rightarrow \infty} x_i(t)= \frac{\sum_{j \in \mathcal{R}_i} x_j(0)}{|\mathcal{R}_i|}
$$
for all $i\in \mathcal{V}$, where $\mathcal{R}_i \subseteq \mathcal{V}$ denotes the set of nodes of the connected component in which node $i$ lies.
\end{definition}

\medskip

 It is well known that  the system (\ref{r5}) reaching componentwise  consensus is equivalent to \cite{Magnus}
$$
\lim_{t \rightarrow \infty} \big\|X(t)\big\|_{L_\mathsf{G}}=0,
$$
where $\big\|X(t)\big\|_{ L_\mathsf{G}}= X^T(t)L_\mathsf{G} X(t)$.
On the other hand, we have from Lemma \ref{lem0} that
$$
\Big\{{\rm \bf vec}(z):  \mathscr{P}_\ast (z)  =z, z\in \mathbb{C}^{ 2^n \times 2^n}\Big\}= {\rm ker} (L_\mathsf{G}).
$$
As a result, the following conclusion holds providing a direct relation between  quantum consensus  and its classical analogue.

\medskip

\begin{theorem}\label{propequivalence}
Quantum consensus  over $\mathsf{G}$ along (\ref{r4}) is equivalent to   componentwise  consensus  in the classical sense
 over the induced graph $\mathcal{G}$ along (\ref{r5}).
\end{theorem}

\medskip

\begin{figure*}[t]
\begin{center}
\includegraphics[height=2.4in]{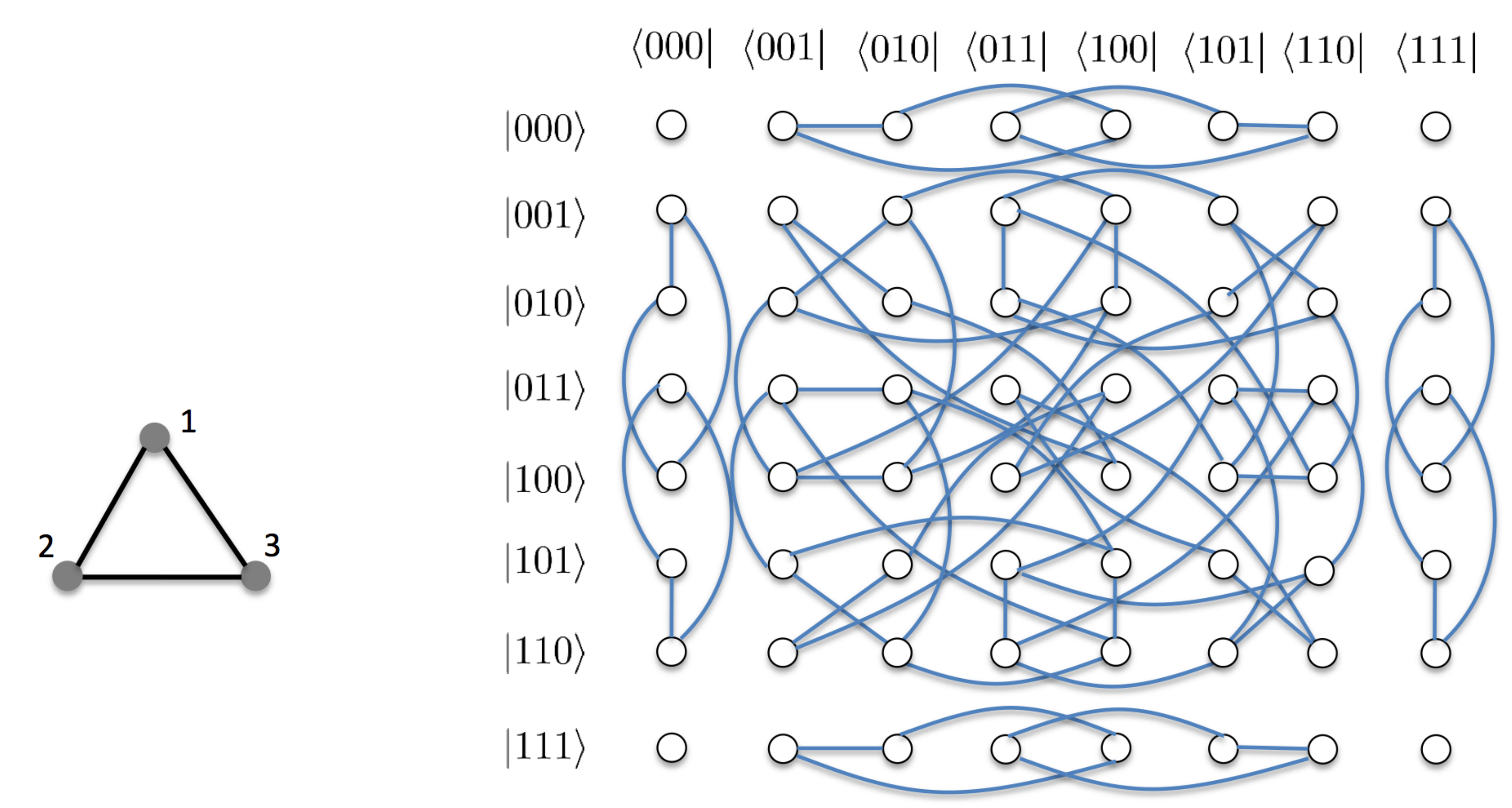}
\caption{The induced graph of the three-qubit quantum complete graph. There are $64$ nodes in the induced graph, and they can be indexed as the elements in the basis $\mathbb{B}$.}
\label{inducegraph}
\end{center}
\end{figure*}

\begin{figure*}[t]
\begin{center}
\includegraphics[height=1.2in]{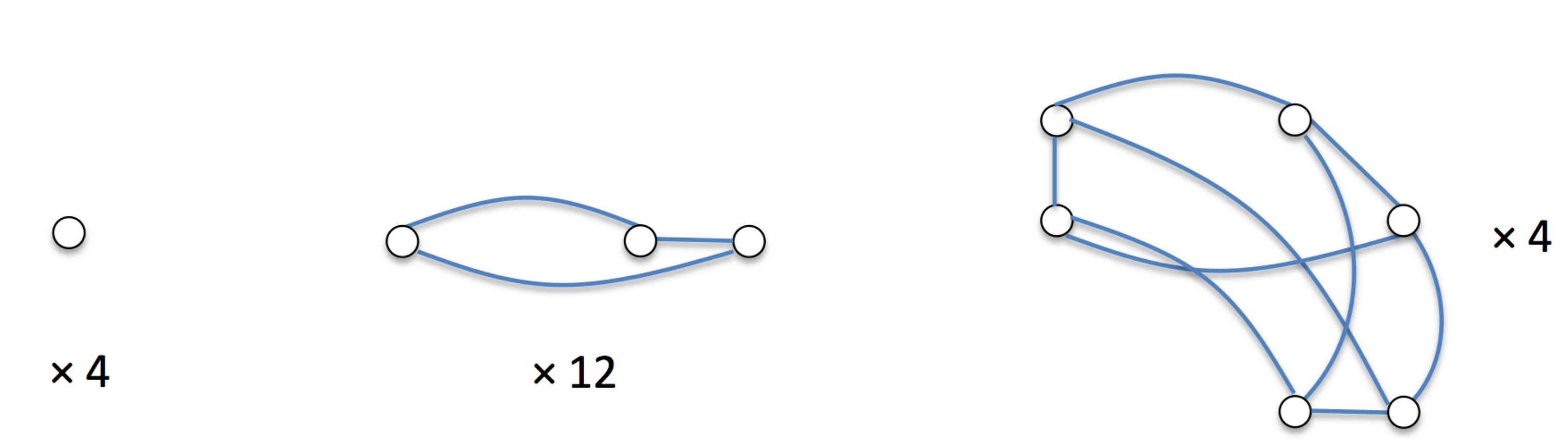}
\caption{The connected components of the induced graph for the three-qubits quantum complete graph. There are a total of $20$ components, consisting of $4$ components each with one node, $12$ components each with three nodes, and the remaining  $4$ components each with six nodes. Note that all of these components are {\it regular} graphs in the sense that every node within the same component has the same degree. } \label{component}
\end{center}
\end{figure*}

\begin{remark}
Theorem \ref{propequivalence} describes a form of quantum parallelism (cf., Chapter 1.4.2, \cite{Nielsen})
in the sense that the original quantum consensus dynamics over $n$ qubits, leads to
independent  consensus processes over  disjoint  subsets of nodes. As shown in Figures \ref{inducegraph} and \ref{component},
if the quantum interaction  graph is well chosen,
the state evolution can be of the same form  for these different subsets of nodes, but starting from (in general) different initial values.
\end{remark}

\subsection{The Connected Components}
We have seen from Theorem \ref{propequivalence} that we can indeed investigate the connected components of the quantum induced graph
$\mathcal{G}$ to obtain every detail  of the quantum consensus master equation.  Now we establish some basic properties of the connected
 components of the quantum induced graph.

\subsubsection{The Reachable Nodes}
We index the elements $\mathcal{V}=\{1,\dots, 2^{2n}\}$ under the standard computational basis of $\mathcal{H}^{\otimes n}$.
Recall that $|0\rangle$ and $|1\rangle$ form a  basis of $\mathcal{H}$.  Let $|q_1\rangle\otimes \dots \otimes |q_n\rangle \in \mathcal{H}^{\otimes n}$
be denoted as $|q_1 \dots q_n\rangle$ for simplicity, where $\otimes$ represents the tensor product. Then, the  following $2^n$ elements
$$
|q_1 \dots q_n\rangle: q_i\in\{0,1\}, i=1,\dots,n
$$
form a basis of $\mathcal{H}^{\otimes n}$.
We define
$$
|q_1 \dots q_n\rangle \langle p_1 \dots p_n|:\mathcal{H}^{\otimes n} \mapsto \mathcal{H}^{\otimes n}
$$
as a linear operator over $\mathcal{H}^{\otimes n}$ such that
$$
\Big( |q_1 \dots q_n\rangle \langle p_1 \dots p_n|\Big) |\xi \rangle = \Big( \langle p_1 \dots p_n  |\xi\rangle \Big) |q_1 \dots q_n\rangle,
$$
for all $|\xi\rangle\in \mathcal{H}^{\otimes n}$.
We now obtain a basis for all linear operators over $\mathcal{H}^{\otimes n}$ (which is  isomorphic to $\mathbb{C}^{ 2^n \times 2^n}$):
$$
\mathbb{B}:=\Big\{|q_1 \dots q_n\rangle \langle p_1 \dots p_n|:  q_i, p_i\in\{0,1\}, i=1,\dots,n \Big\}.
$$

Furthermore,   associated with any $\pi\in \mathbf{P}$ with $\mathbf{P}$ being the permutation group over $\mathsf{V}$, we define an operator $\mathscr{F}_{\pi}$
over $\mathcal{H}^{\otimes n}\times \mathcal{H}^{\otimes n}$ by
$$
\mathscr{F}_{\pi} \Big(|q_1 \dots q_n\rangle \langle p_1 \dots p_n| \Big)=|q_{\pi(1)} \dots q_{\pi(n)}\rangle \langle p_{\pi(1)} \dots p_{\pi(n)}|
$$
for all $|q_1 \dots q_n\rangle \langle p_1 \dots p_n|\in \mathbb{B}$.
Particularly, when $\pi\in \mathbf{P}$ defines a swapping permutation $\pi_{jk}$, the corresponding $\mathscr{F}_{\pi}$ will be denoted as
$\mathscr{F}_{\pi_{jk}}$. Then the following lemma holds with its proof given in Appendix B.

\medskip

\begin{lemma} \label{lemfpi}
For all $\rho \in \mathbb{C}^{ 2^n \times 2^n}$ and $\pi \in \mathbf{P}$, it holds that $U_{\pi}\rho U_{\pi}^\dag=\mathscr{F}_{\pi} (\rho)$.
\end{lemma}

\medskip

Each node in $\mathcal{V}$ corresponds to one entry in $\rho \in \mathbb{C}^{2^n \times 2^n}$ under vectorization. We identify the
nodes in $\mathcal{V}$ as the elements in $\mathbb{B}$. For any $|q_1 \dots q_n\rangle \langle p_1 \dots p_n| \in \mathcal{V}$,
we denote by $\mathcal{N}_{|q_1 \dots q_n\rangle \langle p_1 \dots p_n|}$ the set of nodes in $\mathcal{V}$ that are adjacent
to $|q_1 \dots q_n\rangle \langle p_1 \dots p_n|$ in the induced graph $\mathcal{G}$. It is then clear from Lemma \ref{lemfpi} that
\begin{align*}
&\mathcal{N}_{|q_1 \dots q_n\rangle \langle p_1 \dots p_n|}=\Big\{ |q_{\pi_{jk}(1)} \dots q_{\pi_{jk}(n)}\rangle \langle p_{\pi_{jk}(1)} \dots p_{\pi_{jk}(n)}| \nonumber\\
&\neq |q_1 \dots q_n\rangle \langle p_1 \dots p_n|:\ \  \pi_{jk} \in \mathsf{E}\Big\}.
\end{align*}
Noting that all the swapping permutations in $$
\Big\{\pi_{jk}: \{j,k\}\in \mathsf{E} \Big\}
$$ form a generating subset of $\mathbf{P}$, the following lemma holds.

\medskip

\begin{lemma}\label{lemreachnode}
Suppose $\mathsf{G}$ is connected. Then for any given node $|q_1 \dots q_n\rangle \langle p_1 \dots p_n| \in \mathcal{V}$,
$$
\mathcal{R}_{|q_1 \dots q_n\rangle \langle p_1 \dots p_n|}:=\Big\{ |q_{\pi(1)} \dots q_{\pi(n)}\rangle \langle p_{\pi(1)} \dots p_{\pi(n)}|:\pi\in \mathbf{P} \Big\}
$$
is the set of nodes in $\mathcal{V}$ that are reachable from $|q_1 \dots q_n\rangle \langle p_1 \dots p_n|$ in the graph $ \mathcal{G}$.
\end{lemma}

\subsubsection{Several Counting Theorems}

We now establish some scaling properties of the components of the induced graph. First of all the following theorem  holds, with a detailed proof in Appendix C.

\medskip

\begin{theorem}\label{induceproperty}
Suppose $\mathsf{G}$ is connected. Then
\begin{itemize}
\item [(i).] There are ${\rm dim} \Big( \big\{{\rm \bf vec}(z):  \mathscr{P}_\ast (z)  =z, z\in \mathbb{C}^{ 2^n \times 2^n} \big\} \Big)
$ connected components in  $\mathcal{G}$. Different choices of  $\mathsf{G}$ give the same   node set partition of $\mathcal{V}$ along
the connected components of their induced graphs.

\item [(ii).] Let $|\cdot|$ stand for the cardinality of a finite set. The degree of   $|q_1 \dots q_n\rangle \langle p_1 \dots p_n| \in \mathcal{V}$ is computed as
$\big| \mathcal{N}_{|q_1 \dots q_n\rangle \langle p_1 \dots p_n|}\big|$.

\item [(iii).] There are exactly four smallest components of $\mathcal{G}$, each of which contains only one node.
The number of nodes in the largest components of $\mathcal{G}$ lies in the interval
    $$
\Big[ \max_{0\leq k \leq n} \mathbf{C}_n^k, \Big(\max_{0\leq k \leq n} \mathbf{C}_n^k\Big)^2 \Big]
$$  where  $\mathbf{C}_n^k$ is the combinatorial number of selecting $k$ different elements out of $n$ different choices.
\end{itemize}
\end{theorem}

\medskip

\begin{remark}\label{remarkcoponent}
Note that $ \max_{0\leq k \leq n} \mathbf{C}_n^k$ is achieved at $k=\lfloor \frac{n+1}{2}\rfloor$, where $\lfloor b \rfloor$
denotes the greatest integer no larger than $b$ for a given $b\in\mathbb{R}$.  Invoking the famous Stirling's formula it is known that
$$
\max_{0\leq k \leq n} \mathbf{C}_n^k \sim \frac{2^n}{\sqrt{\pi n/2}}.
$$
Therefore, based on Theorem \ref{induceproperty}, we know that the size of the largest component, asymptotically (as $n$ tends to infinity) lies in
$$
\bigg[  \frac{2^n}{\sqrt{\pi n/2}},\ \   \frac{4^n}{\pi n/2} \bigg].
$$
\end{remark}

Let $\mathbf{K}_n$ denote the complete graph with $n$ nodes. The following theorem  establishes some tight bounds of the node degree for the induced graph, whose proof is  in Appendix D.

\medskip

\begin{theorem}\label{theoremdegree}
(i). If $n \mod 4=0$, then ${\rm deg}(v)\leq 3n^2/8$ for all $v\in\mathcal{V}$;

(ii). If $n \mod 4=1$, then ${\rm deg}(v)\leq (3n^2-3)/8$ for all $v\in\mathcal{V}$;

(iii).  If $n \mod 4=2$, then ${\rm deg}(v)\leq (3n^2-4)/8$ for all $v\in\mathcal{V}$;

(iv). If $n \mod 4=3$, then ${\rm deg}(v)\leq (3n^2-3)/8$ for all $v\in\mathcal{V}$.

Moreover, there exist  nodes with degrees at these upper bounds when $\mathsf{G}=\mathbf{K}_n$.
\end{theorem}

\medskip

\begin{remark}
Theorem \ref{theoremdegree} indicates  that the maximum degree of the induced graph asymptotically tends to
$
{3n^2}/{8}
$ as $n$ tends to infinity.   While the maximum component is of the size at least   ${2^n}/{\sqrt{\pi n/2}}$ from Remark \ref{remarkcoponent}. As a result, the largest components of the induced graph  tend to be rather {\it sparse} as $n$ becomes large.
\end{remark}
\subsubsection{Component Structure}
We now investigate the structure of  the components.  We focus on the case when the quantum interaction graph is the complete graph.

Recall that an undirected graph is {\it regular} if all nodes in the graph have the same degree \cite{godsil}.
We further introduce the following definition \cite{spectrabook}.

\medskip

\begin{definition}
Let $\mathrm{G}$ be a simple, undirected regular graph with $N$ nodes and node degree $k$.  We call $\mathrm{G}$ strongly regular if there are two integers $\lambda$ and $\mu$ such that

(i) Every two adjacent nodes have $\lambda$ neighbors in common;

(ii) Every two non-adjacent nodes have $\mu$ neighbors in common.
\end{definition}

\medskip

We also introduce  the quantum induced graph on the diagonal entries as a subgraph of $\mathcal{G}$.

\medskip

\begin{definition}
The quantum diagonal induced  graph, denoted $\mathcal{G}_{\rm diag}=(\mathcal{V}_{\rm diag}, \mathcal{E}_{\rm diag})$,
 is the subgraph generated by the node set $\mathcal{V}_{\rm diag}:=\big\{|p_1 \dots p_n\rangle \langle p_1 \dots p_n|: \ p_i\in\{0,1\}\big\}$ in the graph $\mathcal{V}$.
\end{definition}

\medskip

With Lemma \ref{lemreachnode}, there are no edges between   $\mathcal{V}_{\rm diag}$ and $\mathcal{V}\setminus \mathcal{V}_{\rm diag}$
in the graph $\mathcal{G}$. The quantum diagonal induced graph $\mathcal{G}_{\rm diag}$ therefore fully characterizes the dynamics  of the diagonal entries of the density operator.
The physical interpretation of the diagonal entries is that $$
[\rho]_{|p_1\dots p_n\rangle \langle p_1\dots p_n |}
$$ represents the probability of finding the system at the state $|p_1\dots p_n\rangle \langle p_1\dots p_n |$
when performing measurement to the quantum network  under the standard basis \cite{Nielsen}.

 The following theorem provides  a structural   characterization of the induced graph. The proof can be found in Appendix E.

 \medskip

\begin{theorem}\label{theoremregular}
Suppose $\mathsf{G}=\mathbf{K}_n$. Then

\noindent (i). Every connected component of the induced graph $\mathcal{G}$ is regular;

\noindent (ii).  Every connected component of the diagonal  induced graph $\mathcal{G}_{\rm diag}$ is almost strongly regular in the sense that
\begin{itemize}
\item[a)] every two adjacent nodes in $\mathcal{G}_{\rm diag}$ have $n-2$ neighbors in common;
\item [b)]  every two non-adjacent nodes in $\mathcal{G}_{\rm diag}$ have either zero or one neighbor in common.
\end{itemize}
\end{theorem}

\medskip

\begin{remark}
 The exponentially increasing dimension with respect to the number of components is a fundamental obstacle for understanding and analyzing large-scale quantum systems. Theorems \ref{propequivalence}, \ref{induceproperty}, \ref{theoremdegree}, and \ref{theoremregular}  illustrate  the  possibility of splitting the dimensions  into decoupled smaller pieces (e.g., Remark \ref{remarkcoponent}, the dimension is reduced by a factor which is at least $2/{\pi n}$) by graphical analysis, and then combinatorial analysis would be able  to uncover deeper characterizations.  The nature of quantum systems engineered by sparse Lindblad operators,  or quantum systems with sparse Hamiltonians, suggests  potential applicability of the methodology to more studies  of quantum multi-body systems \cite{many-body1,manybody2}.
\end{remark}

\subsection{Discussions}
%

\subsubsection{Why Swapping Operators?}
We now provide a brief discussion to illustrate that  the choice of swapping operators in the quantum consensus dynamics (\ref{sys10}),
is very natural from classical consensus dynamics \cite{saber04}. A group-theoretic point of view for their relationships is also provided in \cite{Ticozzi}.

Consider a classical graph $\mathrm{G}=(\mathrm{V}, \mathrm{E})$ with $\mathrm{V}=\{1,\dots,N\}$. Let $x_i(t) \in \mathbb{R}$
 be the state of node $i$ in $\mathrm{V}$. Denote $x(t)=\big(x_1(t)\  \dots\  x_N(t)\big)^T$. Let every edge's weight be one, and let $L_\mathrm{G}$ be the  Laplacian
  in the classical sense of the graph $\mathrm{G}$. Then a classical average consensus process is defined by \cite{jad03,saber04}
\begin{align}\label{105}
\frac{d}{dt}x(t)=- L_\mathrm{G} x(t).
\end{align}

We introduce a classical swapping operator (matrix) along the edge $\{i,j\}\in\mathrm{E}$, denoted by $\widetilde{U}_{ij}\in \mathbb{R}^{N\times N}$, in the way that
\begin{align}
\widetilde{U}_{ij} (z_1 \dots z_i \dots z_j \dots z_N)^T =(z_1 \dots z_j \dots z_i \dots z_N)^T,
\end{align}
for all $(z_1 \dots z_N)^T \in \mathbb{R}^m$.
Then physically $\widetilde{U}_{ij}$ switches the $i$'th and $j$'th entries with the rest unchanged, and is therefore a classical version of
the quantum swapping ${U}_{ij}$. In fact  $\widetilde{U}_{ij}$ is a permutation matrix. It is interesting to note the following equality:
\begin{align}\label{106}
L_{\mathrm{G}}=-\sum_{\{i,j\}\in \mathrm{E}} \Big( \widetilde{U}_{ij}- I_N \Big).
\end{align}
Plugging (\ref{106}) into (\ref{105}), we obtain the following equivalent form of (\ref{105}):
\begin{align}\label{108}
\frac{d}{dt}x(t)= \sum_{\{i,j\}\in \mathrm{E}} \Big( \widetilde{U}_{ij}x(t)-  x(t) \Big).
\end{align}

It is now  clear that the system (\ref{sys10}) is a formal  quantum version of the system (\ref{108}), noting that in the quantum case the
swapping operator $U_{ij}$ maps a density operator $\rho$ to $U_{ij}\rho U_{ij}^\dag$. This is to say, the connection between the quantum
consensus and its classical prototype, is  inherent within  their structures, and the realization of quantum consensus seeking via swapping operators is remarkably  natural.

\begin{remark}
As a matter of fact, the quantum consensus state, defined in (\ref{quantumconsensus}) (originally  introduced in \cite{Ticozzi}),
is formally of the same form as the classical average noticing
\begin{align}\label{classicalconsensus}
\frac{1}{N!} \sum_{\pi \in \mathbf{P}} \widetilde{U}_\pi z= \frac{1}{N!}\cdot \big((N-1)! \big){\bf 1}_{N}^T z {\bf 1}_N= \frac{\sum_{i=1}^n z_i}{N} {\bf 1}_N
\end{align}
for all $z=(z_1 \cdots z_N)^T \in \mathbb{C}^{N}$, where $ \widetilde{U}_\pi$ denotes the classical permutation.
 We have now seen that the classical average (\ref{classicalconsensus}) and the quantum average   (\ref{quantumconsensus}) are closely connected.
\end{remark}

\subsubsection{Convergence Speed Optimization}
  If each edge $\{i,j\}\in \mathsf{E}$ is associated with a weight  $\alpha_{ij}$, we can correspondingly define the weighted quantum Laplacian  $
L_\mathsf{G}(\alpha):=\sum_{\{j,k\}\in \mathsf{E}} \alpha_{jk} \big(I_{2^n}\otimes I_{2^n}-  U_{jk} \otimes U_{jk }  \big)$ with $\alpha=(\alpha_{jk}: \{j,k\}\in \mathsf{E})$.

The speed of convergence to a quantum consensus for
\begin{align}
\frac{d \rho}{dt}=\sum_{\{j,k\}\in \mathsf{E}}\alpha_{jk} \Big(U_{jk}\rho U_{jk}^\dag  -\rho\Big),
\end{align}
is thus given by the smallest non-zero eigenvalue of $L_\mathsf{G}(\alpha)$, denoted $\lambda_2\big(L_\mathsf{G}(\alpha)\big)$.

As a continuous-time and quantum analogue of \cite{fast}, we can therefore  optimally distribute a certain amount, say $W_0>0$,
of edge weights onto the edges so that the fastest convergence rate can be achieved:
\begin{equation}\label{eq3}
\begin{aligned}
& {\text{maximize}}
& & \lambda_2\Big(L_\mathsf{G}(\alpha)\Big) \\
& \text{subject to}
& & \sum_{\{i,k\} \in \mathsf{E}}\alpha_{jk} \leq W_0.
\end{aligned}
\end{equation}
Following similar argument as in \cite{fast}, we know that $\lambda_2\big(L_\mathsf{G}(\alpha)\big)$ is a concave function of $\alpha$.
Therefore, the fastest convergence can be obtained by solving (\ref{eq3}) via standard  convex programming methods.

\medskip

We conclude this section with a few remarks. In this section we have provided a graphical approach for studying the quantum  consensus
master equation. We introduce the quantum
Laplacian and the quantum induced graph, and show that quantum consensus over the interaction graph is equivalent to
 componentwise classical consensus over the induced graph, with convergence rate given by the smallest eigenvalue of the quantum Laplacian.
 We establish some basic  properties of the induced graph in terms its scaling and structure. Such a fundamental connection makes the majority of  graphical developments in classical
 network systems directly applicable to  quantum networks.  The proposed graphical approach  certainly also  applies  to  discrete-time quantum dynamics, e.g., \cite{Ticozzi}.

\medskip

\section{Quantum Synchronization}\label{Secsynchronization}
In this section, we establish synchronization conditions for  the Lindblad equation  (\ref{sysLind}). First of all, making use of the graphical approach developed in the previous section,
we establish two necessary and sufficient quantum consensus conditions for  the system (\ref{sys10}) in light of existing results on classical consensus. Next, we show that for a class
of network Hamiltonians, quantum consensus of the system (\ref{sys10}) implies synchronization of the system (\ref{sysLind}). Finally,
we discuss the connection between the quantum synchronization
results and their classical analogue and present a numerical example.

\subsection{Quantum Consensus Conditions}
The following theorem establishes consensus conditions of the system (\ref{sys10}).

\medskip

\begin{theorem}\label{theoremconsensus}(i) The system (\ref{sys10}) achieves  global  exponential  quantum consensus  if and only if there exists a constant $T>0$
such that $\mathsf{G}([t,t+T)):=(\mathsf{V},\bigcup_{t\in[t,t+T)} \mathsf{E}_{\sigma(t)} )$ is connected for all $t\geq0$.

(ii) The system (\ref{sys10}) achieves global  asymptotic quantum consensus if and only if  $\mathsf{G}([t,\infty)):=(\mathsf{V},\bigcup_{t\in[t,\infty)} \mathsf{E}_{\sigma(t)} )$
is connected for all $t\geq0$.
\end{theorem}

\medskip

The proof of Theorem \ref{theoremconsensus} is based on the connection between  quantum consensus and classical consensus from a graphical point of view, and has been put in Appendix F.
These results are essentially consistent with the results for consensus seeking over classical networks \cite{tsi}-\cite{shisiam}.
   We remark that under the conditions of Theorems \ref{theoremconsensus},
   the convergence rates can be explicitly computed making use of the analysis in \cite{shisiam},
    for both cases. We also remark that for simplicity of presentation we assume the edge weights $\alpha_{jk}$ to be a constant.
    Generalization to the case where $\alpha_{jk}$ is time-varying  or even state-dependent is straightforward using  existing works
    in the literature on classical consensus convergence, e.g., \cite{shisiam}.

\begin{remark}
Theorem \ref{theoremconsensus} provides a generalization to the result in \cite{Ticozzi-MTNS} for switching quantum interaction graphs. In fact, from its proof
it is clear that the convergence rate can be obtained utilizing the results in \cite{shisiam} under the given conditions.
\end{remark}

\subsection{From Consensus to Synchronization}
Let the initial time be $t_0 =0$ and denote $\rho_\ast=\mathscr{P}_\ast (\rho(0))$.  Introduce $$
\tilde{\rho}(t) = e^{\imath Ht/\hbar} \rho(t) e^{-\imath H t/\hbar}.
$$ Suppose  $[H, U_\pi]=0$ for all $\pi\in\mathbf{P}$.
Then some simple calculations lead to the fact that the evolution of $\tilde{\rho}(t)$ satisfies
\begin{align}\label{sys2}
\frac{d \tilde{\rho}}{dt}=\sum_{\{j,k\}\in \mathsf{E}} \alpha_{jk}  \Big(U_{jk}\tilde{\rho} U_{jk}^\dag  -\tilde{\rho}\Big).
\end{align}
Substituting  the results in Theorem \ref{theoremconsensus},  we immediately obtain
\begin{align}\label{100}
\lim_{t\to \infty} \Big[\rho(t)-e^{-\imath Ht/\hbar} \rho_\ast e^{\imath H t/\hbar} \Big]=0
\end{align}
when the same connectivity conditions hold in Theorem \ref{theoremconsensus} for the switching quantum interaction graph.

 Define $\rho^k_\ast(t):={\rm Tr}_{\otimes_{j\neq k} \mathcal{H}_j } \Big(e^{-\imath Ht/\hbar} \rho_\ast e^{\imath H t/\hbar}\Big)$ for all $k\in \mathsf{V}$.
  The following lemma can be established from the definition of the partial trace \cite{Nielsen} (or, directly applying Theorem 1 in \cite{Ticozzi}).

  \medskip

 \begin{lemma}
 Suppose $[H, U_\pi]=0$ for all $\pi\in\mathbf{P}$.  Then $\rho^k_\ast(t)=\rho^m_\ast(t)$ for all $k,m\in \mathsf{V}$ and all $t$.
\end{lemma}

\medskip

As a result, the following theorem holds.

\medskip

\begin{theorem}
\label{theoremsynchronization} Suppose $[H, U_\pi]=0$ for all $\pi\in\mathbf{P}$.

\noindent (i). If $\mathsf{G}([t,\infty)):=(\mathsf{V},\bigcup_{t\in[t,\infty)} \mathsf{E}_{\sigma(t)} )$ is connected for all $t\geq0$,
then  the system (\ref{sysLind}) achieves global
asymptotical quantum  (reduced-state) synchronization.

\noindent (ii). If there exists a constant $T>0$ such that $\mathsf{G}([t,t+T)):=(\mathsf{V},\bigcup_{t\in[t,t+T)} \mathsf{E}_{\sigma(t)} )$ is connected for all $t\geq0$,
then the system (\ref{sysLind}) achieves global exponential quantum  (reduced-state) synchronization.
\end{theorem}

\medskip

The following lemma, with its proof given in Appendix G, presents two  classes of Hamiltonians satisfying the condition $[H, U_\pi]=0$ for all $\pi\in\mathbf{P}$.
Denote the Kronecker sum $H_0^{\oplus n}=\sum_{i=1}^n I^{\otimes (i-1)}\otimes H_0 \otimes I^{\otimes (n-i)}$,  where $H_0$ is  a Hermitian operator over $\mathcal{H}$.

\medskip

\begin{lemma}\label{lem8}
Let $H_0$ be  a Hermitian operator over $\mathcal{H}$. If either  $H= H_0^{\otimes n}$ or $H=H_0^{\oplus n}$ holds, then  $[H, U_\pi]=0$ for all $\pi\in\mathbf{P}$.
\end{lemma}

\medskip

\begin{remark}
If  $H= H_0^{\oplus n}$, then there holds  $e^{\imath Ht/\hbar}=  e^{\imath H_0t/\hbar} \otimes \cdots \otimes  e^{\imath H_0t/\hbar} $ and
$e^{-\imath Ht/\hbar}=  e^{-\imath H_0t/\hbar} \otimes \cdots \otimes  e^{-\imath H_0t/\hbar} $. Consequently, it can be further deduced that
\begin{align}\label{102}
\rho^k_\ast(t)&={\rm Tr}_{\otimes_{j\neq k} \mathcal{H}_j } \Big(e^{-\imath Ht/\hbar} \rho_\ast e^{\imath H t/\hbar}\Big)\nonumber\\
&=e^{-\imath H_0t/\hbar} \Big( {\rm Tr}_{\otimes_{j=1}^{n-1} \mathcal{H}_j } \big(\rho_\ast \big) \Big)  e^{\imath H_0 t/\hbar}.
\end{align}
from the definition of the partial trace \cite{Nielsen}.
\end{remark}

\subsection{Discussions}
It is worth noticing that the quantum synchronization  results established in Theorem \ref{theoremsynchronization}, is exactly the quantum analogues  of the classical studies on
 the synchronization of coupled oscillators \cite{WU-CHUA95,PRL1998,WUbook}. Fundamental results have been derived for the classical notion of synchronization  for the following
 dynamics \cite{WU-CHUA95,PRL1998,WUbook}:
\begin{align}\label{classicalsync}
\frac{d}{dt}x_i(t)=Ax_i(t)+\sum_{j=1}^N W_{ij} \big(x_j(t)-x_i(t) \big), \ \ i=1,\dots, N
\end{align}
where $x_i\in \mathbb{R}^m$, $A\in\mathbb{R}^{m\times m}$, $W_{ij}\geq 0$. Here $x_i(t)$ represents the state of the $i$'th oscillator, $A$ is the inherent mode of the dynamics
of the oscillators, and an interaction graph is induced by $[W_{ij}]$. Note that it is critical that all of the oscillators share an identical
 inherent dynamics  for synchronization of the system (\ref{classicalsync}). Therefore, it becomes clear that the condition $H=H_0^{\oplus n}$ plays the same role in imposing  identical inherent dynamics for the qubits.  The system~(\ref{sysLind})
 becomes the quantum equivalence   of the system~(\ref{classicalsync}) when such a condition holds, and the behavior of the system trajectories in the two systems are indeed
 consistent \cite{WU-CHUA95}. On the other hand, for the case with $H= H_0^{\otimes n}$, the tensor product of Hamiltonians introduces internal interactions   among the qubits.
 Synchronization of the qubits' reduced states is still reached since these internal interactions {\it cooperate} with the (external)  swapping interactions in such a way that
  $H$ is invariant under permutations. It is however difficult to write down the explicit trajectory of each qubit's reduced state as a function of $H_0$ in this case, and
  the synchronization orbit  is certainly no longer  the one determined   by  $H_0$ for the most choices of $H_0$.

\begin{remark}
Note that when the nodes' inherent self-dynamics are not identical in the classical synchronization dynamics (\ref{classicalsync}), it is
well-known in the literature that it will be extremely difficult and often impossible to
achieve synchronization for the system (\ref{classicalsync}) \cite{WUbook}.
Now that it becomes clear from above discussion that the condition that either  $H= H_0^{\otimes n}$ or $H=H_0^{\oplus n}$
in the quantum master equation plays the same  role in enforcing
identical inherent self-dynamics,  quantum synchronization will in general be difficult to reach without such conditions.
\end{remark}

\subsection{Numerical Example}
In this subsection, we present a simple numerical example to illustrate the above quantum synchronization result.

We consider three qubits indexed in $\mathsf{V}=\{1,2,3\}$. Their interaction graph is  fixed as the complete graph, i.e.,
 $\mathsf{E}=\big\{\{1,2\}, \{2,3\}, \{1,3\} \big\}$.
Let $\alpha_{12}=\alpha_{13}=\alpha_{23}=1$. The initial network state is chosen to be
$$
\rho_0=\frac{1}{2}|10 0\rangle \langle 100|+\frac{1}{2}|10 0\rangle \langle 101|+\frac{1}{2}|10 1\rangle \langle 101|+\frac{1}{2}|10 1\rangle \langle 100|.
$$
The network Hamiltonian is chosen to be $H=\sigma_z\otimes \sigma_z\otimes \sigma_z$, where
\begin{equation}
\sigma_z=\left(\begin{matrix}
1 & 0\\
0 & -1\\
\end{matrix}\right)
\end{equation}
is  one of the Pauli matrices.

\begin{figure}[t]
\begin{center}
\includegraphics[height=2.5in]{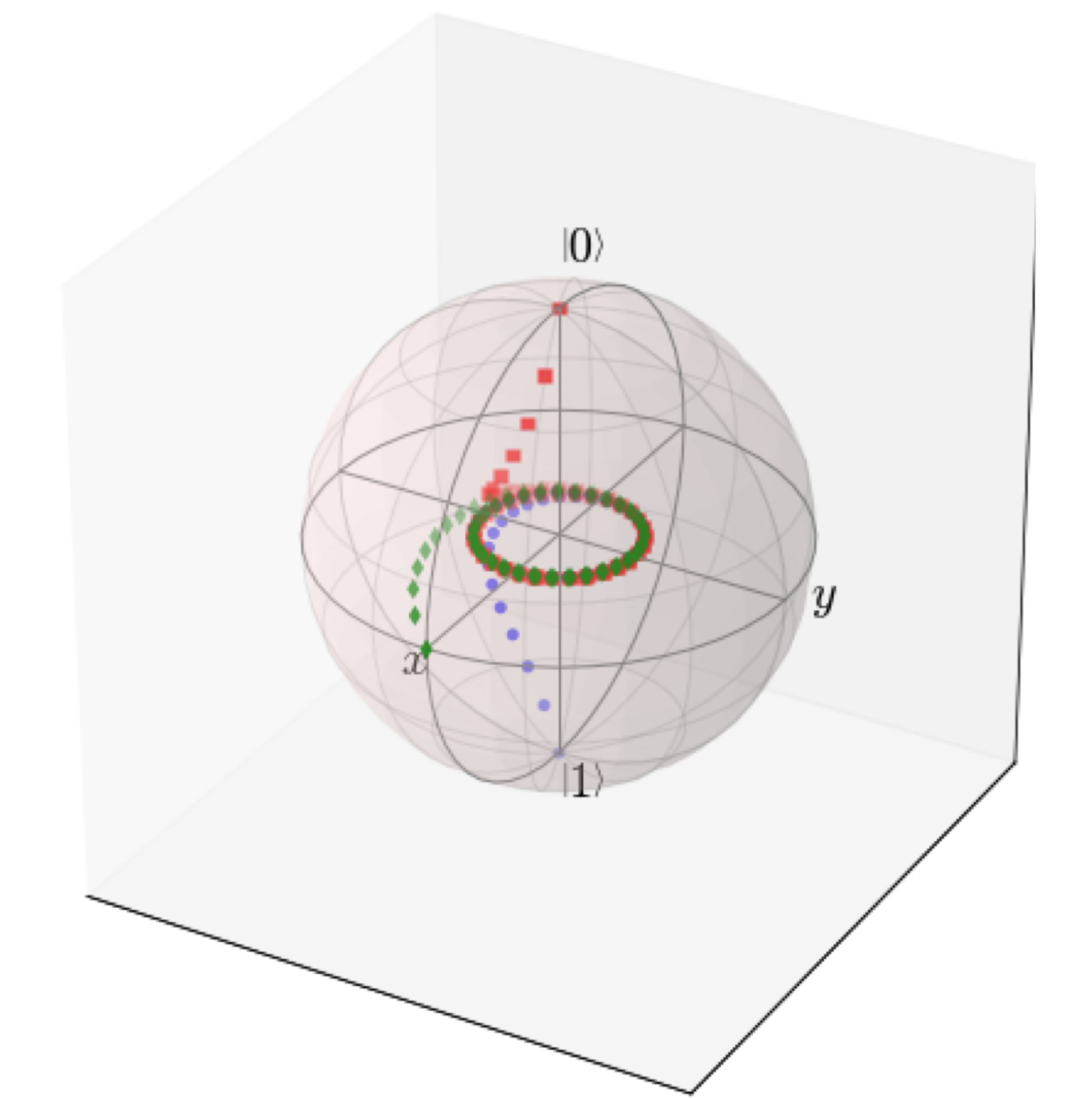}
\caption{An illustration of the quantum synchronization: The orbits of the three qubits asymptotically converge to  the same trajectory for the proposed master equation. } \label{blochspere}
\end{center}
\end{figure}

We first plot the evolution of the reduced states of the three qubits on one Bloch sphere. Clearly their orbits
 asymptotically tend to the same trajectory determined  by
the Hamiltonian  $\sigma_z$ (cf., Figure \ref{blochspere}).

\begin{figure}[t]
\begin{center}
\includegraphics[height=2.5in]{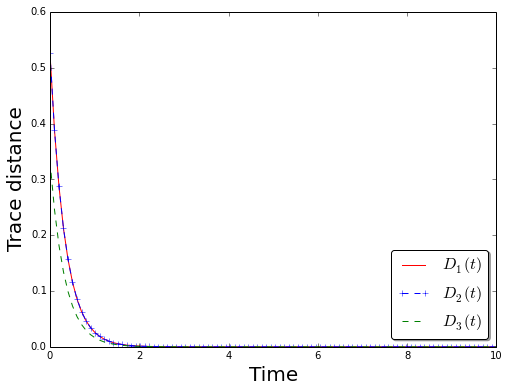}
\caption{An illustration of the quantum synchronization: Exponential convergence to the synchronization orbit.
Note that two of the three qubits' distance functions exactly agree with each other so there are only two curves distinguishable in this plot.  } \label{figure_reduce_convergence}
\end{center}
\end{figure}

Next, recall that the trace distance between two density operator $\rho_1$, $\rho_2$ over the same Hilbert space,
denoted by $\|\rho_1-\rho_2\|_{\rm Tr}$, is defined as
$$
\Big\|\rho_1-\rho_2\Big\|_{\rm Tr}=\frac{1}{2}{\rm Tr}\sqrt{\Big(\rho_1-\rho_2\Big)^\dag\Big(\rho_1-\rho_2\Big)}.
$$
We then plot the trace distances between the reduced states and the synchronization orbit,
$$
D_k(t):=\Big\|\rho^k (t)- {\rm Tr}_{\otimes_{j=1}^{2} \mathcal{H}_j } \Big( e^{-\imath H t/\hbar}  \rho_\ast  e^{\imath H t/\hbar}   \Big)\Big\|_{\rm Tr}
$$
for $ k=1,2,3$, as a function of $t$, where  $\rho_\ast=\frac{1}{3!} \sum_{\pi \in \mathbf{P}_3} U_\pi \rho_0 U^\dag_\pi$ is
the quantum average with $\mathbf{P}_3$ denoting the permutation group with order three. Clearly they all
converge to zero with an exponential rate (cf., Figure \ref{figure_reduce_convergence}).

\section{Conclusions}\label{Sec6}
We have investigated consensus and synchronization problems for a quantum network with $n$ qubits. The state evolution of the quantum network equipped with
continuous-time swapping operators,  is  described by a Lindblad master equation.   These swapping operators also introduce  an underlying interaction graph.
 A graphical method bridging  the proposed quantum consensus scheme and  classical consensus dynamics was presented,
  by studying an induced graph (with $2^{2n}$ nodes) of the  quantum interaction graph (with $n$ qubits).
  We provided  several fundamental relations between a quantum graph and its induced classical graph.
  Two necessary and sufficient conditions for exponential and asymptotic quantum consensus were obtained, respectively,
   for switching quantum interaction graphs. We also presented  quantum synchronization conditions, in the sense that
    the reduced states of all qubits tend to a common trajectory. We showed that this is exactly the quantum analogue of classical synchronization of coupled oscillators.

The consensus and synchronization problems for the quantum network  considered
 in this paper can be taken as a special class of stabilization problems in quantum control \cite{Altafini and Ticozzi 2012}-\cite{Qi and Guo 2010}
  where the control actions are realized by swapping operators. We believe the results presented in the current paper add some novel understandings
   regarding the control and state manipulation of quantum networks in a distributed manner. The graphical approach proposed may   serve as a systematic and useful
    tool for analyzing distributed quantum dynamics. In future, it is also worth investigating new  algorithms for other consensus/synchronization  states in quantum networks and
    developing control methods for stabilizing the states of quantum networks.

\section*{Appendices}

\subsection*{Appendix A. Proof of Lemma \ref{lem0}}
The following equalities hold:
  \begin{align} \label{r1}
{\rm ker} ({L}_\mathsf{G})&=\Big\{{\rm \bf vec}(z): \sum_{\{j,k\}\in \mathsf{E}} \big(U_{jk}z U_{jk}^\dag -z\big)=0\Big\}\nonumber\\
  &\stackrel{a)}{=} \Big\{{\rm \bf vec}(z): U_{jk}z U_{jk}^\dag =z, \{j,k\}\in \mathsf{E}\Big\} \nonumber\\
  &\stackrel{b)}{=} \Big\{{\rm \bf vec}(z): U_\pi z U_\pi^\dag =z, \pi\in \mathbf{P}\Big\} \nonumber\\
  &\stackrel{c)}{=} \Big\{{\rm \bf vec}(z):  \mathscr{P}_\ast (z)  =z\Big\}.
  \end{align}
Here $a$) is based on Lemma 5.2 in \cite{PRAinformation}; $b$) holds from the fact that $\mathsf{G}$ is a connected graph so that the swapping permutations along the edges among qubits consist of a generating set of the group $\mathbf{P}$ (cf. Proposition 8 and Lemma 1 of \cite{Ticozzi}). Regarding equality $c$), on one hand it is straightforward to see that $$
\Big\{{\rm \bf vec}(z): U_\pi z U_\pi^\dag =z, \pi\in \mathbf{P}\Big\} \subseteq \Big\{{\rm \bf vec}(z):  \mathscr{P}_\ast (z)  =z\Big\}.
 $$
 On the other hand, if  $\mathscr{P}_\ast (z)  =z$, then
$$
U_\pi z U_\pi^\dag=U_\pi  \big( \mathscr{P}_\ast (z) \big)  U_\pi^\dag=\mathscr{P}_\ast (z)=z
$$
since $\pi \mathbf{P}=\mathbf{P}$ for any $\pi\in \mathbf{P}$. Thus we also have
$$
\Big\{{\rm \bf vec}(z):  \mathscr{P}_\ast (z)  =z\Big\} \subseteq \Big\{{\rm \bf vec}(z): U_\pi z U_\pi^\dag =z, \pi\in \mathbf{P}\Big\}.
$$
 This proves  the desired lemma.  \hfill$\square$

\subsection*{Appendix B. Proof of Lemma \ref{lemfpi}}
 Since the two operators:
$$
\rho \rightarrow U_{\pi}\rho U_{\pi}^\dag
$$
and
$$
\rho \rightarrow \mathscr{F}_{\pi} (\rho)
$$
are both linear, we just need to verify the equality for each element in the basis $\mathbb{B}$.

The following holds:
\begin{align}
&\Big( U_{\pi} |q_1 \dots q_n\rangle \langle p_1 \dots p_n| U_{\pi}^\dag \Big) |\xi\rangle\nonumber\\
 &=  \langle p_1 \dots p_n|U_{\pi}^\dag |\xi\rangle U_{\pi} |q_1 \dots q_n\rangle \nonumber\\
&= \Big(\langle   p_{\pi(1)} \dots p_{\pi(n)} |\xi\rangle \Big) |q_{\pi(1)}  \dots q_{\pi(n)} \rangle \nonumber\\
&= \Big( |q_{\pi(1)}  \dots q_{\pi(n)} \rangle  \langle p_{\pi(1)} \dots p_{\pi(n)} | \Big)|\xi\rangle \nonumber\\
&=  \mathscr{F}_{\pi} \Big( |q_1 \dots q_n\rangle \langle p_1 \dots p_n|  \Big) |\xi\rangle
\end{align}
for any $|\xi\rangle \in \mathcal{H}^{\otimes n}$. This proves the desired lemma. \hfill$\square$

\subsection*{Appendix C. Proof of Theorem \ref{induceproperty}}
(i). The number of connected components of $\mathcal{G}$ has been derived in Lemma \ref{lemnumber}. The fact that the sizes of  $\mathcal{G}$'s connected components
do not depend on the form of $\mathsf{G}$, as long as $\mathsf{G}$ is connected,  can be simply deduced from Lemma \ref{lemreachnode}.

\noindent (ii). The conclusion holds directly from the proof of Lemma \ref{lemreachnode}.

\noindent (iii). First of all note that the following four nodes
$
|0\dots 0\rangle\langle 0\dots 0|$,  $|0\dots 0\rangle\langle 1\dots 1|$,  $|1\dots 1\rangle\langle 0\dots 0|$,  $|1\dots 1\rangle\langle 1\dots 1|$
are always isolated in $\mathcal{G}$ since both  $|0\dots 0\rangle$ and $|1\dots 1\rangle$ are invariant under any permutation $\pi \in \mathbf{P}$.
Furthermore, it is easy to see that for a node
$$
|q_1 \dots q_n\rangle \langle p_1 \dots p_n| \in\mathcal{V}
$$
to be isolated, it must be the case that both $ |q_1 \dots q_n\rangle$ and $|p_1 \dots p_n\rangle$ are invariant  under any permutation $\pi \in \mathbf{P}$.
This proves that the four isolated nodes presented above are the only four isolated nodes in $\mathcal{G}$.

Finally, we establish the upper and lower bounds  to the number of nodes in the largest component. The following claim holds.

\medskip

\noindent {\it Claim.} $\big| \big\{ |q_{\pi(1)} \dots q_{\pi(n)}\rangle, \pi\in \mathbf{P} \big\} \big|=\mathbf{C}_n^r$ with $r=\sum_{k=1}^{n} q_k$.

\medskip

For any $|q_{1} \dots q_{n}\rangle$ and  $|p_{1} \dots p_{n}\rangle$ with $\sum_{k=1}^{n} q_k=\sum_{k=1}^{n} p_k$, we can always find a permutation $\pi^\sharp \in \mathbf{P}$
such that $|q_{1} \dots q_{n}\rangle= |p_{\pi^\sharp(1)} \dots p_{\pi^\sharp(n)}\rangle$. As a result, $ \big\{ |q_{\pi(1)} \dots q_{\pi(n)}\rangle, \pi\in \mathbf{P} \big\}$ has
$\mathbf{C}_n^r$ elements. This proves the claim.

From Lemma \ref{lemreachnode},  as long as either $|q_1 \dots q_n\rangle\neq  |q_{\pi(1)} \dots q_{\pi(n)}\rangle $ or $|p_1 \dots p_n\rangle\neq  |p_{\pi(1)} \dots p_{\pi(n)}\rangle $ holds,
$\pi$ will generate a reachable node for $|q_1 \dots q_n\rangle \langle p_1 \dots p_n|$.
Then the upper and lower bounds for the size of the largest component in $\mathcal{G}$ follows immediately.

The proof is now complete. \hfill$\square$
\subsection*{Appendix D. Proof of Theorem \ref{theoremdegree} }
The argument is based on a  combinatorics analysis on the choice of nodes under the basis $\mathbb{B}$. We present the detailed proof for Cases (i) and (iii).
The remaining two cases can be  proved via the same techniques, and whose details  are therefore omitted.

\noindent (i). Let $n=2m$ with some positive integer $m\geq 1$ and take a node $v\in\mathcal{V}$. Without loss of generality, we assume $v$ takes the form
$$
\big|\underbrace{0\dots 0}_{2\chi} \underbrace{1\dots 1}_{2m-2\chi} \big\rangle \big \langle p_1 \dots p_{2m}\big|,
$$
where $p_j\in \{0,1\}$ and $0\leq \chi \leq m$. It is clear that a quantum link $\{j,k\}\in \mathsf{E}$ (i.e., operator $\pi_{jk}$) generates a neighbor of
 node $v$ only for the following three cases:
\begin{itemize}
\item[a).]  $j\leq \chi$ and $k\geq \chi+1$, or   $k\leq \chi$ and $j\geq \chi+1$;
\item[b).] $j\leq \chi$ and $k\leq\chi$ with $p_j \neq p_k$;
\item[c).] $ j\geq \chi+1$ and $k\geq \chi+1$ with $p_j \neq p_k$.
\end{itemize}
Consequently,  direct combinatorial calculations lead to
\begin{align}
{\rm deg}(v)&\leq \chi^2+(m-\chi)^2+2\chi(2m-2\chi)\nonumber\\
&=-2\chi^2+2m\chi+m^2\nonumber \\
&\leq \frac{3m^2}{2}.
\end{align}
Moreover, the upper  bound ${3m^2}/{2}$ is reached when $\mathsf{G}=\mathbf{K}_n$, $m$ is even (i.e., $n \mod 4=0$), and $v$ is of the form with $\chi=m/2$:
$$
\big|\underbrace{0\dots 0}_{2\chi} \underbrace{1\dots 1}_{2m-2\chi} \big\rangle \big \langle \underbrace{0\dots 0}_{\chi} \underbrace{1\dots 1}_{\chi} \underbrace{0\dots 0}_{m-\chi} \underbrace{1\dots 1}_{m-\chi} \big|.
$$
This proves (i).

\noindent (iii). Again let $n=2m$ with some positive integer $m\geq 1$. We study the case when $v$ takes the form
$$
\big|\underbrace{0\dots 0}_{2\chi+1} \underbrace{1\dots 1}_{2m-2\chi-1} \big\rangle \big \langle p_1 \dots p_{2m}\big|,
$$
where $p_j\in \{0,1\}$ and $2\chi+1 \leq 2m$. Via similar analysis we have
\begin{align}
{\rm deg}(v)&\leq \chi(\chi+1)+(m-\chi)(m-\chi-1)\nonumber\\
&+(2\chi+1)(2m-2\chi-1)\nonumber\\
&=-2\chi^2+2(m-1)\chi+m^2+m-1\nonumber \\
&\leq \frac{3m^2-1}{2}.
\end{align}
The upper  bound ${(3m^2-1)}/{2}$ is reached when $\mathsf{G}=\mathbf{K}_n$, $m$ is odd  (i.e., $n \mod 4=2$), and $v$ is of the form with $\chi=(m-1)/2$:
$$
\big|\underbrace{0\dots 0}_{2\chi+1} \underbrace{1\dots 1}_{2m-2\chi-1} \big\rangle \big \langle \underbrace{0\dots 0}_{\chi} \underbrace{1\dots 1}_{\chi+1} \underbrace{0\dots 0}_{m-\chi} \underbrace{1\dots 1}_{m-\chi-1} \big|.
$$
This proves (iii).
 \hfill$\square$

\subsection*{Appendix E. Proof of Theorem \ref{theoremregular}}
(i). Let $|q_1 \dots q_n\rangle \langle p_1 \dots p_n|$ and $|q'_1 \dots q'_n\rangle \langle p'_1 \dots p'_n|$ be two nodes in $\mathcal{V}$ belonging
 to a common    component, where $q_i,p_i,q'_i,q'_i$ take values from $\{0,1\}$. From Lemma \ref{lemreachnode}, we know that we can find a permutation
 $\pi_\ast \in\mathbf{P}$ such that
\begin{align}
|q'_1 \dots q'_n\rangle \langle p'_1 \dots p'_n|=|q_{\pi_\ast(1)} \dots q_{\pi_\ast (n)}\rangle \langle p_{\pi_\ast(1)} \dots p_{\pi_\ast(n)}|.
\end{align}

Now suppose $\pi_{jk}$ generates a link to node $|q_1 \dots q_n\rangle \langle p_1 \dots p_n|$ in the induced graph, i.e., $|q_1 \dots q_n\rangle \langle p_1 \dots p_n|\neq   |q_{\pi_{jk}(1)} \dots q_{\pi_{jk}(n)}\rangle \langle p_{\pi_{jk}(1)} \dots p_{\pi_{jk}(n)}|$. We define a swapping permutation $\pi^\natural$ by
$$
\pi^\natural=\pi_{\pi_\ast(j) \pi_\ast(k)}.
$$
In other words, $\pi^\natural$ flips the state of qubits $\pi_\ast(j)$ and $ \pi_\ast(k)$. This gives us
\begin{align}
&|q'_1 \dots q'_n\rangle \langle p'_1 \dots p'_n|\nonumber\\
&=|q_{\pi_\ast(1)} \dots q_{\pi_\ast (n)}\rangle \langle p_{\pi_\ast(1)} \dots p_{\pi_\ast(n)}|\nonumber\\
&\neq  |q_{\pi^\natural\pi_\ast(1)} \dots q_{\pi^\natural\pi_\ast (n)}\rangle \langle p_{\pi^\natural\pi_\ast(1)} \dots p_{\pi^\natural\pi_\ast(n)}|\nonumber\\
&=|q'_{\pi^\natural(1)} \dots q'_{\pi^\natural(n)}\rangle \langle p'_{\pi^\natural(1)}\dots p'_{\pi^\natural(n)}|.
\end{align}
Consequently, $\pi^\natural$, as an edge in $\mathsf{G}$ since $\mathsf{G}=\mathbf{K}_n$, also generates a link to node $|q'_1 \dots q'_n\rangle \langle p'_1 \dots p'_n|$
 in the induced graph. Noting that the positions of $|q_1 \dots q_n\rangle \langle p_1 \dots p_n|$ and $|q'_1 \dots q'_n\rangle \langle p'_1 \dots p'_n|$ are symmetric
  in the above argument, we have constructed a bijection between the adjacent nodes  of  $|q_1 \dots q_n\rangle \langle p_1 \dots p_n|$ and those of
   $|q'_1 \dots q'_n\rangle \langle p'_1 \dots p'_n|$. This proves the desired conclusion.

\noindent (ii). From the proof of Theorem \ref{induceproperty} we know that
\begin{align}\label{r6}
\mathcal{R}_{|p_1 \dots p_n\rangle \langle p_1 \dots p_n|}=\Big\{ |p'_1 \dots p'_n\rangle \langle p'_1 \dots p'_n| :\sum_{k=1}^n p'_k= \sum_{k=1}^n p_k \Big\}.
\end{align}
For two nodes $v=|p_1 \dots p_n\rangle \langle p_1 \dots p_n|$ and $v'=|p'_1 \dots p'_n\rangle \langle p'_1 \dots p'_n|$ in the same component of the  diagonal
induced graph, we introduce
$$
H(v,v')=\sum_{k=1}^n\big |p_k-p'_k\big|.
$$

\medskip

\noindent Proof of Condition a): let $v=|p_1 \dots p_n\rangle \langle p_1 \dots p_n|$ and $v'=|p'_1 \dots p'_n\rangle \langle p'_1 \dots p'_n|$ be two adjacent nodes in the diagonal induced graph.
 As a result, we have  $H(v,v')=2$ and $\sum_{k=1}^n p'_k= \sum_{k=1}^n p_k=L$ for some integer $L\leq n$. The following claim holds.

\medskip

{\it Claim.} There are $n-2$ common neighbors  for $v$ and $v'$.

\medskip

Since $H(v,v')=2$, without loss of generality, we write $v=|01p_3 \dots p_n \rangle \langle01p_3 \dots p_n|$ and $v'=|10p_3 \dots p_n \rangle \langle01p_3 \dots p_n|$.
If $p_3=0$, then it is straightforward to see  that
$$
|001p_4\dots p_n \rangle \langle001p_4 \dots p_n|
$$
is a common neighbor of $v$ and $v'$. Similarly if $p_3=1$,  a common neighbor of $v$ and $v'$ is given as
$$
|110p_4\dots p_n \rangle \langle001p_4 \dots p_n|.
$$
Continuing  the argument to $p_4,\dots,p_n$ we can find $n-2$ common neighbors for $v$ and $v'$.  Apart from these $n-2$ common neighbors,
either  $v$ or $v'$ however has only two more neighbors as themselves. This proves the claim.

\medskip

\noindent Proof of Condition b):   let $v=|p_1 \dots p_n\rangle \langle p_1 \dots p_n|$ and $v'=|p'_1 \dots p'_n\rangle \langle p'_1 \dots p'_n|$ be two non-adjacent nodes in the same component.
This means that $H(v,v')>2$. From (\ref{r6}) we know that $H(v,v')$ must be an even number. Thus, $H(v,v')\geq 4$.
 On the other hand, let $v^\flat :=|p^\flat_1 \dots p^\flat_n\rangle \langle p^\flat_1 \dots p^\flat_n|$ be a common neighbor of $v$ and $v'$.
 Then $H(v,v^\flat)=2$ and $H(v',v^\flat)=2$, which yields $H(v,v')\leq 4$. Consequently,
 we can easily conclude that $v$ and $v'$ have exactly one common neighbor if $H(v,v')=4$,
 and they have no common neighbor if $H(v,v')>4$.

The proof is now complete. \hfill$\square$

\subsection*{Appendix F. Proof of Theorem \ref{theoremconsensus}}

The proof is based on the graphical approach developed in Section \ref{SecGraphical}. Under vectorization, the system (\ref{sys10})
 is equivalent to the following vector form:
 \begin{align}\label{sys11}
 \frac{d}{dt} {\bf vec}({\rho}(t)) =-L({\sigma(t)}){\bf vec}({\rho}(t)),
 \end{align}
where by definition
$$
L({\sigma(t)}):=\sum_{\{j,k\}\in \mathsf{E}_{\sigma(t)}}\alpha_{jk}\Big(I_{2^n}\otimes I_{2^n}-  U_{jk} \otimes U_{jk }  \Big).
$$

We denote the induced graph of the quantum interaction graph
$\mathsf{G}_{\sigma(t)}= (\mathsf{V},\mathsf{E}_{\sigma(t)} )$, as $\mathcal{G}_{\sigma(t)}=\big(\mathcal{V},\mathcal{E}_{\sigma(t)}\big)$. The following lemmas hold.

\medskip

\begin{lemma}\label{lem2}
Let $T>0$ be a constant. Then $\mathcal{G}([t,t+T))$ has  $m^\natural={\rm dim} \big( \big\{{\rm \bf vec}(z):  \mathscr{P}_\ast (z)  =z\big\} \big)$
connected components if $\mathsf{G}([t,t+T))$ is connected. \end{lemma}
{\it Proof.}  Noticing the fact that $\mathcal{G}([t,t+T))$ is the induced graph of $\mathsf{G}([t,t+T))$ following Definition \ref{definitioninduce},
the desired lemma holds directly from Lemma \ref{lemnumber}.  \hfill$\square$

\medskip

\begin{lemma}\label{lem6}
Suppose $\mathsf{G}([0,\infty))$ is connected. Then the system (\ref{sys11}) defines $m^\natural$ classical consensus processes over
$m^\natural$ disjoint subsets of nodes in $\mathcal{V}$.
\end{lemma}
{\it Proof.}  We will  show it using Lemma \ref{lem2}.
If $\mathsf{G}([0,\infty))$ is connected, then $\mathcal{G}([0,\infty))$ has  $m^\natural$ connected components. This means that for
 any two nodes belonging to different connected components of  $\mathcal{G}([0,\infty))$, there is never an edge  between them for the system (\ref{sys11}).
 This implies the desired conclusion.  \hfill$\square$

\medskip

We now denote the $m^\natural$ disjoint subsets of nodes in $\mathcal{V}$, each defining the node set of one component of
 $\mathcal{G}([0,\infty))$ when $\mathsf{G}([0,\infty))$ is connected, as $\mathcal{V}_1,\dots,\mathcal{V}_{m^\natural}$. Correspondingly, we denote by  $$
\mathcal{G}_{\sigma(t)}^o= (\mathcal{V}_o,\mathcal{E}_{\sigma(t)}^o), \ o=1,\dots,m^\natural
$$
the subgraph that is associated with  $\mathcal{V}_o$ in the graph $\mathcal{G}_{\sigma(t)}$. We give another technical lemma.

\medskip

\begin{lemma}\label{lem7} Suppose $\mathsf{G}([0,\infty))$ is connected. Then
\begin{itemize}
\item [(i)] The system (\ref{sys10}) reaches global (exponential, or asymptotic) quantum consensus if and only if the system (\ref{sys11})
reaches classical global (exponential or asymptotic) consensus over all node subsets $\mathcal{V}_o,  o=1,\dots,m^\natural$.
    \item [(ii)] Let $T>0$ be a constant. Then $\mathcal{G}^o([t,t+T)):=(\mathcal{V}_o,\bigcup_{t\in[t,t+T)} \mathcal{E}_{\sigma(t)}^o)$
     is connected for all $o=1,\dots, m^\natural$ if and only if $\mathsf{G}([t,t+T))$ is connected.
\end{itemize}
\end{lemma}
{\it Proof.} (i). First of all we fix the initial time as $t_0=0$ and the initial value for $\rho(0)$, and show the equivalence between quantum
consensus and classical consensus. The fact  that classical consensus is reached  for the system (\ref{sys11}) means that
$$
\lim_{t\rightarrow \infty} x_i(t)= \frac{\sum_{j \in \mathcal{V}_o} x_j(0)}{|\mathcal{V}_o|},\ \ i\in  \mathcal{V}_o,\ o=1,\dots,m^\natural
$$
where again we use the  notation $X(t)=(x_1 (t) \dots x_{4^n}(t))^T:= {\bf vec}(\rho(t))$, since each $L({\sigma(t)})$ is always symmetric. This in turn implies that
$$
\lim_{t \rightarrow \infty} \big\|X(t)\big\|_{L_\mathsf{G}}=0
$$
for an arbitrary connected $\mathsf{G}$. Thus, quantum consensus is equivalent to classical consensus for this fixed initial condition.

Next, it is clear  that $\rho(0)$ taking value from all legitimate density operators  makes $X_o(0)=(x_k(0): k\in \mathcal{V}_o)^T$ possibly take value from a unit ball in
 $\mathbb{R}^{|\mathcal{V}_o|}$. This implies that global quantum consensus for the system (\ref{sys10}) is equivalent to global consensus for  the system (\ref{sys11}).

Finally, the convergence rate equivalence (exponential, or asymptotic), is obvious since $m^\natural$ defines a finite number.

(ii). Noticing the definition of connected component and Lemma \ref{lem2}, the desired conclusion follows immediately.  \hfill$\square$

It is straightforward to see that  $\mathsf{G}([0,\infty))$ must be connected so that
quantum consensus convergence becomes possible for the $n$-qubit network. Based on Theorem 4.1 in \cite{shisiam},   global
exponential consensus is achieved for the component $\mathcal{V}_o$  if and only if there exists
$T>0$ such that $\mathcal{G}^o([t,t+T))$ is connected for all $t$.  Theorem 5.2 in  \cite{shisiam} showed  that  global asymptotic  consensus is achieved for the component $\mathcal{V}_o$ if and only if
$\mathcal{G}^o([t,\infty))$ is connected for all $t$. As a result, utilizing Lemma \ref{lem7}
on the equivalence between quantum consensus and classical consensus,  Theorem \ref{theoremconsensus}
immediately holds. This concludes the proof. \hfill$\square$

\subsection*{Appendix G. Proof of Lemma \ref{lem8}}
We only prove the lemma for case (i) and the other case follows from a similar argument. Take $\pi\in\mathbf{P}$. The following holds:
\begin{align}
&\Big[ H_0\otimes \cdots \otimes H_0\Big]  U_\pi  \Big(|q_1 \dots q_n\rangle \Big)\nonumber\\
&=\big|H _0q_{\pi(1)} \big\rangle \otimes \dots \otimes \big|H_0 q_{\pi(n)}\big\rangle\nonumber\\
&=U_\pi \Big(\big|H_0 q_1\rangle \otimes \dots \otimes \big| H_0q_n\big\rangle\Big)\nonumber\\
&=  U_\pi\Big[ H_0\otimes \cdots \otimes H_0\Big]   \Big(|q_1 \dots q_n\rangle \Big)
\end{align}
for  all $|q_1 \dots q_n\rangle \in \mathcal{H}^{\otimes n}$. This immediately implies  $[H, U_\pi]=0$ and the desired conclusion thus holds. \hfill$\square$

\section*{Acknowledgements}

 This work was supported in part by the Australian Research Council under projects DP130101658 and FL110100020,
the Knut and Alice Wallenberg Foundation, and the Swedish Research Council.
The authors gratefully  thank Dr. Shuangshuang Fu for her  generous help in preparing the numerical example.

\medskip

\medskip

\medskip

\medskip

\noindent {\sc Guodong Shi} \\
\noindent {\small College of Engineering and Computer Science, The Australian National University, \\ Canberra, ACT 0200 Australia}\\  {\small Email: } {\tt\small guodong.shi@anu.edu.au}

\medskip

\medskip

\noindent {\sc Daoyi  Dong and Ian R. Petersen} \\
\noindent {\small School of Engineering and Information Technology, University of New South Wales,\\
Canberra, ACT 2600 Australia}\\  {\small Email: } {\tt\small daoyidong@gmail.com, i.r.petersen@gmail.com}

 \medskip

\medskip
\noindent {\sc Karl H. Johansson} \\
\noindent {\small  ACCESS Linnaeus Centre,
   School of Electrical Engineering,
KTH Royal Institute of Technology,
\\ Stockholm 100 44, Sweden }\\
       {\small Email: 』 {\tt\small kallej@kth.se}

\end{document}